\documentclass[apj, preprint2]{aastex6}

\usepackage{setspace}
\usepackage{graphicx}
\usepackage{upquote,textcomp}
\usepackage{amssymb}
\usepackage{esvect}
\usepackage{subfigure}
\usepackage{hepnames}
\usepackage{float}
\usepackage{multirow}
\usepackage{verbatim}
\usepackage{changepage}
\usepackage{algpseudocode}
\usepackage{algorithm}
\usepackage{newcent}
\usepackage{cases}
\usepackage{hyperref}
\usepackage{setspace}
\usepackage{wasysym}
\usepackage{braket}
\usepackage[utf8]{inputenc}
\usepackage[T1]{fontenc}
\usepackage{todo}

\usepackage{fancyvrb}

\makeatletter
\let \@sverbatim \@verbatim
\def \@verbatim {\@sverbatim \verbatimplus}
{\catcode`'=13 \gdef \verbatimplus{\catcode`'=13 \chardef '=13 }} 
\makeatother 

\makeatletter
\newcommand{\pushright}[1]{\ifmeasuring@#1\else\omit\hfill \displaystyle#1 \fi\ignorespaces}
\newcommand{\pushleft }[1]{\ifmeasuring@#1\else\omit \displaystyle#1 \hfill\fi\ignorespaces}
\newcommand{\tdv}{$\mathrm{T} \delta \mathrm{V}$}
\makeatother

\VerbatimFootnotes
\begin{document}
\title[Determining exoplanetary oblateness using transit depth variations]{Determining exoplanetary oblateness using transit depth variations}
\author{John B. Biersteker\altaffilmark{1} and Hilke Schlichting\altaffilmark{1, 2}}
\altaffiltext{1}{Massachusetts Institute of Technology, 77 Massachusetts Avenue, Cambridge, MA 02139-4307, USA}
\altaffiltext{2}{UCLA, 595 Charles E. Young Drive East, Los Angeles, CA 90095, USA}
\begin{abstract}
The measurement of an exoplanet's oblateness and obliquity provides insights into the planet's internal structure and formation history. Previous work using small differences in the shape of the transit light curve has been moderately successful, but was hampered by the small signal and extreme photometric precision required. The measurement of changes in transit depth, caused by the spin precession of an oblate planet, was proposed as an alternative method. Here, we present the first attempt to measure these changes. Using \textit{Kepler} photometry, we examined the brown dwarf Kepler-39b and the warm Saturn Kepler-427b. We could not reliably constrain the oblateness of Kepler-39b. We find transit depth variations for Kepler-427b at $90.1\%$ significance ($1.65\sigma$) consistent with a precession period of $P_\mathrm{prec} = 5.45^{+0.46}_{-0.37}~\mathrm{years}$ and an oblateness, $f~=~0.19^{+0.32}_{-0.16}$. This oblateness is comparable to Solar System gas giants, and would raise questions about the dynamics and tidal synchronization of Kepler-427b.
\end{abstract}

\maketitle
\section{Introduction}
The determination of the oblateness ($f$) of an exoplanet would shed light on that planet's internal structure, dynamics and formation history.
Specifically, it may be possible to empirically constrain the rotation rate, $P_\mathrm{rot}$, obliquity, $\theta$, and gravitational zonal quadrupole moment, $J_2$, of an exoplanet.

Two methods for determining the oblateness of an exoplanet from transit photometry have been proposed. As a result of the planet being slightly aspherical, an oblate exoplanet's transit light curve differs slightly from that of a perfectly spherical planet with the same cross-sectional area; this effect occurs primarily during the ingress and egress phase of the transit \citep{2002ApJ...572..540H, 2002ApJ...574.1004S, 2003ApJ...588..545B}. 
Efforts to observe this effect have met with mixed results. \citet{2010ApJ...709.1219C} used \textit{Spitzer Space Telescope} photometry to constrain the oblateness of HD 189733b to be less than that of Saturn, the most oblate Solar System planet. A similar search through \textit{Kepler} candidates conducted by \citet{2014ApJ...796...67Z} yielded a tentative detection of oblateness for the ${\sim}20~M_\mathrm{Jup}$ object Kepler-39b (KOI 423.01) and constraints on the oblateness of three other \textit{Kepler} candidates. 
These efforts were hampered by the short duration of the expected signal and its relatively small amplitude, ${\sim} 200 \ \mathrm{ppm}$ for Saturn-like oblateness and ${\sim} 2~\mathrm{ppm}$ for a hot Jupiter \citep{2010ApJ...716..850C}.

A second signal of planetary oblateness was identified by \citet{2010ApJ...709.1219C}. 
If the oblate planet has non-zero obliquity, its spin-axis will precess, and the projected area of the planet, and hence the observed transit depth, will change over the period of that precession. 
For planets as oblate as Jupiter or Saturn, the transit depth may change by ${\sim} 1 \%$, or ${\sim} 100 \ \mathrm{ppm}$ for a Jupiter-like planet around a Sun-like star \citep{2010ApJ...716..850C}. \citeauthor{2010ApJ...716..850C} suggest that this is a more feasible observable, within the precision likely to be achieved by \textit{Kepler}.

In this paper, we present an attempt to observe this signal in \textit{Kepler} photometry. We begin with a summary of the expected signal in Section \ref{sec: oblateness-signal}, discuss our transit depth measurement method in Section \ref{sec: delta measurement}, describe our oblateness detection technique in Section \ref{sec: findoblateness}, present results in Section \ref{sec: results}, and discuss our findings in Section \ref{sec: discussion}.

\section{Planetary Oblateness and the Transit Signal}
\label{sec: oblateness-signal}
Following \citet{2010ApJ...716..850C}, we consider the planet to be an oblate spheroid. The \textit{oblateness}, or \textit{flatness}, is
\begin{align}
\label{eq: f}
f = \frac{R_\mathrm{eq} - R_\mathrm{pol}}{R_\mathrm{eq}} \mathrm{,}
\end{align}
where $R_\mathrm{eq}$ and $R_\mathrm{pol}$ are the equatorial and polar radii, respectively. For rotationally-induced oblateness,
\begin{align}
\label{eq: f rot}
f \approx \frac{3}{2}J_2 + \frac{1}{2} \frac{R_\mathrm{eq}^3}{G M_p} \left( \frac{2 \pi}{P_\mathrm{rot}} \right)^2 \mathrm{,}
\end{align}
where $M_p$ is the planet mass \citep{1999ssd..book.....M}. The \textit{obliquity}, $\theta$, is the angle between the planet's orbit-normal and the spin-axis of the planet. We define $\phi(t)$ as the azimuth of the spin-axis projected onto the orbit-plane. For the case of uniform precession,
\begin{align}
\label{eq: phi}
\phi(t) = \frac{2 \pi (t - t_0)}{P_\mathrm{prec}} \mathrm{,}
\end{align}
where $P_\mathrm{prec}$ is the precession period and $t_0$ is a phase offset. This precession causes the projected area of the exoplanet to change, changing the planet to star area ratio, $\delta(t)$. \citet{2010ApJ...716..850C} derive
\begin{align}
\label{eq: delta}
\delta(t) = k^2 \sqrt{1 - \epsilon^2 \{1 - [\sin{\theta} \cos{\phi} \sin{i} + \cos{\theta} \cos{i}]^2 \}}
\end{align}
for the areal ratio of the planet to the star, where $ k = R_\mathrm{eq} / R_* $ is the planet-to-star radius ratio, $i$ is the transit inclination, and $\epsilon$ is the \textit{ellipticity},
\begin{align}
\label{eq: ellipticity}
\epsilon = \sqrt{1 - (1 - f)^2} \mathrm{.}
\end{align}
From Equation \eqref{eq: delta}, the authors show that the amplitude of the transit depth variations is determined by a combination of oblateness and obliquity:
\begin{align}
\left( \frac{\delta_\mathrm{max}}{\delta_\mathrm{min}} \right)^2 - 1 \approx 2 f \sin^2{\theta} \mathrm{,}
\label{eq: f theta degeneracy}
\end{align}
where the approximation is valid when $f$ is small and $i \approx 90^{\circ}$.
While the amplitude of depth variations can be readily determined from the photometric data, there is a degeneracy between $f$ and $\theta$. 
Since the obliquity determines the extent that the aspect angle of the planet changes over the course of its precession, even very high oblateness can be compensated by low obliquity, producing low amplitude variations. This degeneracy cannot be broken without additional constraints \citep{2010ApJ...716..850C}.

The timescale of the change in transit depth is determined by $P_\mathrm{prec}$ which, for uniform precession in a fixed orbit, is $2 \pi / (\alpha \cos{\theta})$, where
\begin{align}
\alpha = \frac{3}{2} \left( \frac{2 \pi}{P_\mathrm{orb}} \right)^2 \left( \frac{P_\mathrm{rot}}{2 \pi} \right) \frac{J_2}{\lambda}\mathrm{,}
\end{align}
is the precession constant \citep{1975Sci...189..377W, 2004AJ....128.2501W} and $\lambda = I / (M_p R_p^2)$ is the normalized moment of inertia of the planet. Using $\lambda / J_2 = 13.5$, estimated for Saturn \citep{2004AJ....128.2501W}, we obtain the scaling relation
\begin{align}
\label{eq: precession period}
P_\mathrm{prec} = 13.3 \ \mathrm{years} \times \left( \frac{P_\mathrm{orb}}{15 \ \mathrm{d}} \right)^2 \left( \frac{10 \ \mathrm{hr}}{P_\mathrm{rot}} \right) \left( \frac{\lambda / J_2}{13.5} \right) \frac{1}{\cos{\theta}} \mathrm{.}
\end{align}
As this relation shows, the stronger torques from the host star on planets with short periods reduce the precession period. This enables easier detection. 
But, this effect must be balanced against the tidal synchronization of a planet by its host star. If the planet is too close, tidal interaction with the star will slow its rotation period to its orbital period, greatly diminishing any rotationally-induced oblateness. The rate at which a planet's spin, $\omega$, is diminished is
\begin{align}
\frac{d\omega}{dt} = -\frac{9}{4} G M_*^2 \frac{R_\mathrm{eq}^3}{M_p Q_p a^6 \lambda} \mathrm{,}
\end{align}
where $Q_p$ is the planet's tidal dissipation factor, $M_*$ is the stellar mass, $G$ is the gravitational constant, and $a$ is the semimajor axis of the planet's orbit \citep{1966Icar....5..375G}. Integrating this equation with an initial rotation period of $P_{\mathrm{rot},~i}$ yields a spin-down time of
\begin{multline}
\label{eq: spin down time}
\tau_\mathrm{spin} = 1.22 \ \mathrm{Gyr} \times \left( \frac{M_p}{M_\mathrm{Jup}} \right) \left( \frac{Q_p}{10^{6.5}} \right)
\left( \frac{\lambda}{0.25} \right) \\
\times \left( \frac{P_\mathrm{orb}}{15 \ \mathrm{d}} \right)^4 \left( \frac{R_\mathrm{Jup}}{R_\mathrm{eq}} \right)^3 
\left(\frac{10 \ \mathrm{hr}}{P_{\mathrm{rot},~i}} - \frac{10 \ \mathrm{hr}}{P_\mathrm{rot}} \right) \text{.}
\end{multline}
Both the tidal synchronization timescale and precession period are strongly dependent on the orbital period.
\subsection{Candidate Selection}

Based on the above properties of the expected signal, we selected a handful of candidates to scrutinize for evidence of transit depth variations. We began with an expansion of the ``sweet spot'' identified by \citet{2010ApJ...716..850C}. 
\citeauthor{2010ApJ...716..850C} suggested that candidates with $P_\mathrm{orb} = 15 - 30 \ \mathrm{days}$ were likely to have both precession periods observable over \textit{Kepler's} planned six year mission, and spin-down timescales of ${\sim}1~\mathrm{Gyr}$. Because of the shorter actual duration of the primary \textit{Kepler} mission, and the considerable uncertainty in spin-down time estimates, we included planets within a period range of 10-30 days in our search.

To select gas giants, we required $R_p > 6~R_\oplus$. We then restricted our search to confirmed planets only. This was primarily motivated by an estimated false positive rate for \textit{Kepler} giant planet candidates with $P < 400~\mathrm{days}$ of $54.6 \pm 6.5\%$ \citep{2016A&A...587A..64S}. It also allowed for independent determination of stellar and planetary parameters, particularly the mass of the planet.

At the time of the study, only nine planets matched these criteria. To avoid complication from overlapping transits and transit timing variations, we also eliminated the five confirmed multi-planet systems, leaving four candidates. From these four, we selected Kepler-39b (KOI 423.01) and Kepler-427b (KOI 192.01), detailed in Table \ref{table: candidates}.

\begin{deluxetable}{lll}
\tabletypesize{\scriptsize}
\tablewidth{\columnwidth}
\tablecolumns{3}
\tablecaption{Candidate Systems \label{table: candidates}}
\tablehead{
\colhead{~} & \colhead{Kepler-39} & \colhead{Kepler-427}
}
\startdata
Star mass $M_*~\mathrm{[M_\odot]}$ & $ 1.26^{+0.07}_{-0.06} $ & $ 0.96 \pm 0.06 $ \\
Star radius $R_*~\mathrm{[R_\odot]}$ & $ 1.25 \pm 0.03 $ & $ 1.35 \pm 0.20 $ \\
Orbital period $P_{\mathrm{orb}}~\mathrm{[day]}$ & $ 21.087 $ & $ 10.291 $ \\
Semimajor axis $a~\mathrm{[AU]}$ & $ 0.162 \pm 0.003 $ & $ 0.091 \pm 0.010 $ \\
Planet mass $M_p~\mathrm{[M_\mathrm{Jup}]}$ & $ 19.1 \pm 1.0 $ & $ 0.29 \pm 0.09 $ \\
Planet radius $R_p~\mathrm{[R_\mathrm{Jup}]}$ & $ 1.11 \pm 0.03 $ & $ 1.23 \pm 0.21 $ \\
\enddata
\tablecomments{Values for Kepler-39 from the circular orbit model in \citet{2015AA...575A..85B}. Kepler-427 parameters from \citet{2014AA...572A..93H}.}
\end{deluxetable}

\section{Transit Depth Measurements}
\label{sec: delta measurement}
Having selected candidates, we measured the depth of their transits over the course of the \textit{Kepler} mission. When \citeauthor{2010ApJ...716..850C} proposed searching for transit depth variations, the mission was just beginning. They were able to recover the oblateness and other parameters from simulated transit photometry generated with a white noise model ($\sigma = 95 \ \mathrm{ppm}$).
Real \textit{Kepler} photometry, naturally, is a bit messier; in particular, secular trends in photometry and starspots produce confounding signals which could create spurious transit depth variations. We attempted to correct for these without introducing other false signals.

\subsection{Initial Cleaning}
For each candidate, we began with the Pre-search Data Conditioning (PDC-MAP) photometry. Data points with known defects (e.g. reaction wheel desaturation events) were removed. 
Depending on the candidate, this eliminated ${\sim} 10\%$ of the data points. Subsequently, points lying more than $3 \sigma$ from a local linear model were removed. 
This preserved nearly all the data points; fewer than $1\%$ were typically removed. Finally, the PDC-MAP pipeline attenuates long period signals, assuming that they are caused by systematic error. Signals with a period longer than $20$ days are almost entirely removed  \citep{L11thedata}.
Despite this, some quarters (or months, in the case of short cadence photometry) still exhibited an overall slope, especially in data with pronounced ``ramp up'' events. Assuming this long-term trend to be artificial, we de-trended each quarter (or month) using a degree 2 polynomial.

\subsection{Light Curve Normalization with Starspots}
\label{sec: norm with spots}
\begin{figure*}[t]
	\centering
	\includegraphics{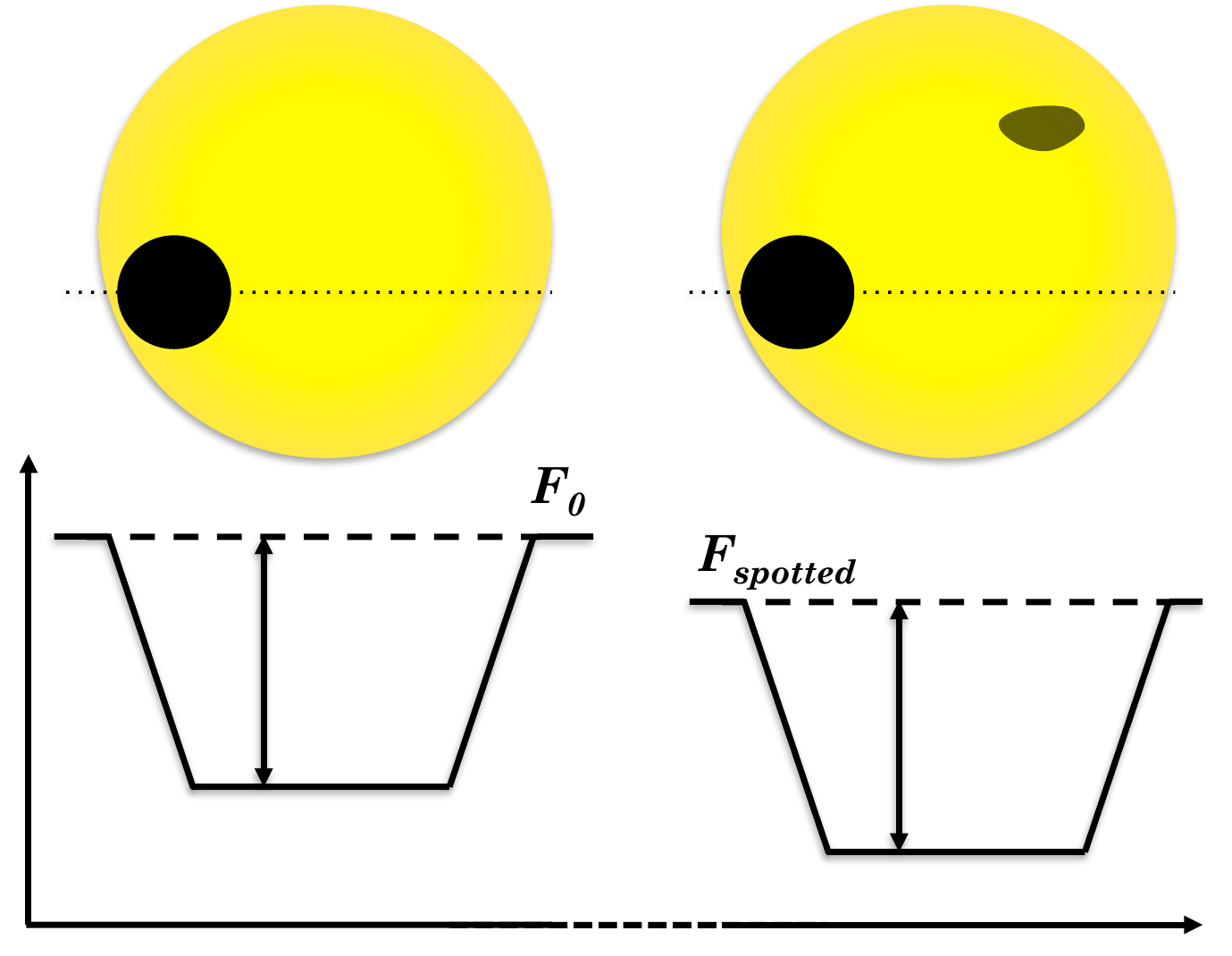}
	\caption{\small An unspotted transit and, some time later, a transit with starspots. The total drop in flux, $\Delta F$, is unchanged, so normalizing by the out-of-transit flux produces a change in the planet-to-star radius ratio.}
	\label{fig: starspots}
\end{figure*}

Cool starspots on the disk of the star produce quasi-periodic variations in stellar flux which can complicate the interpretation of transit light curves \citep[see][]{2009A&A...505.1277C, 2011ApJ...730...82C, 2013ApJ...775...54S}.
As illustrated in Figure \ref{fig: starspots}, starspots outside the transit chord reduce the total flux from the star. But since the flux blocked by the planet is unchanged, the relative transit depth is increased. \citet{2009A&A...505.1277C} found that correcting for this can change the transit depth by ${\sim}1\%$\textemdash comparable to the expected signal from oblateness.

\begin{figure}[ht]
	\centering
	\includegraphics[width=0.45\textwidth]{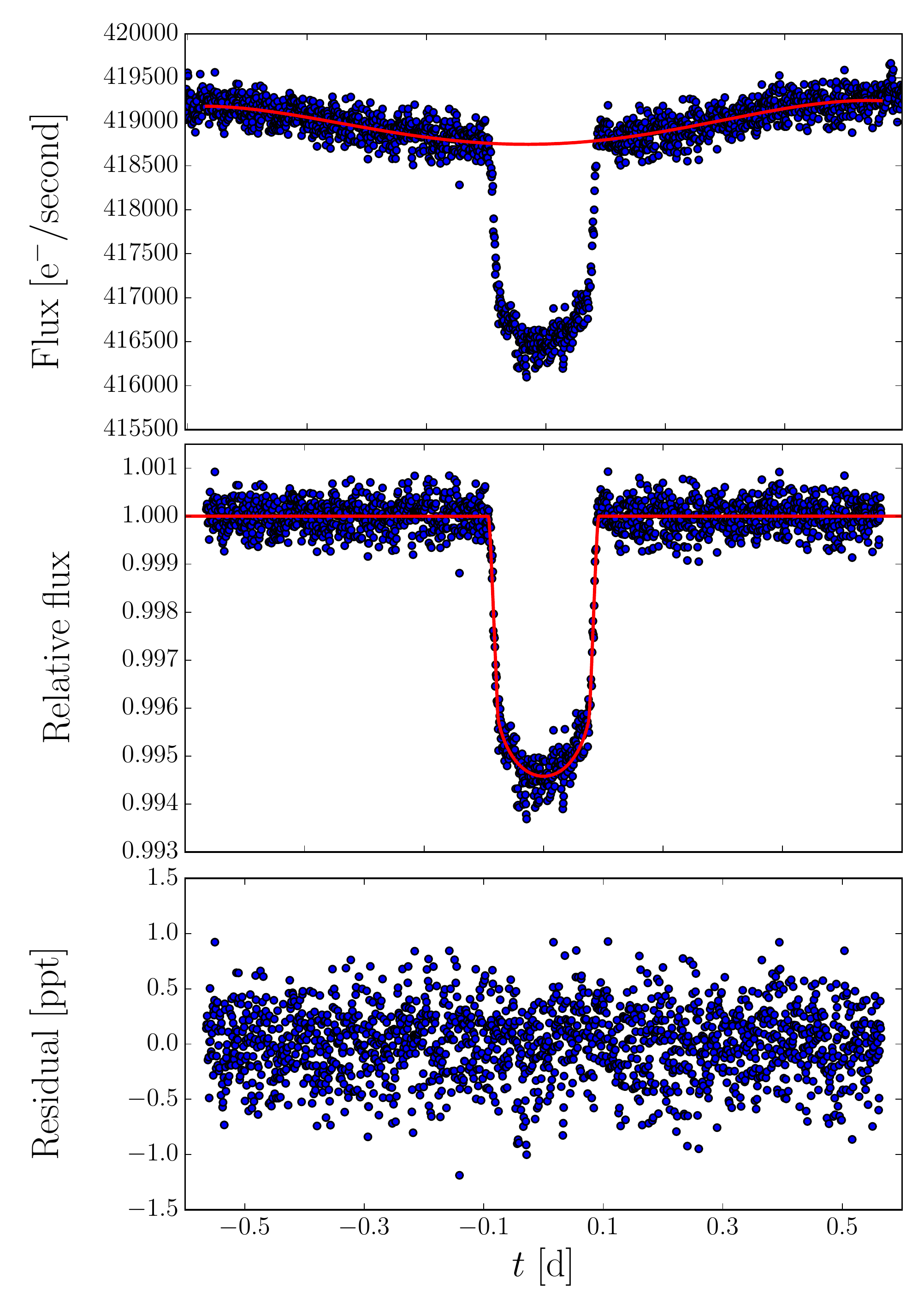}
	\caption{\small Normalization of short cadence photometry from KIC~5812701 and fit to an injected transit of a Saturn-like planet. \textbf{Top:} Raw photometry with the injected transit and stellar flux polynomial model in red. \textbf{Middle:} Normalized transit light curve with quadratic limb darkening model in red. \textbf{Bottom:} Resulting model fit residuals.}
	\label{fig: lc normalization}
\end{figure}

We follow the method outlined by \citeauthor{2009A&A...505.1277C}, normalizing the difference between the measured flux and estimated stellar flux by the estimated ``unspotted'' flux:
\begin{align}
F_{\mathrm{norm},~i} = 1 + \frac{F_i - F_{\mathrm{star},~i}}{F_\mathrm{unspotted}}\mathrm{,}
\end{align}
where $F_i$ is the measured flux value at a time $t_i$, $F_{\mathrm{star},~i}$ is the estimated stellar flux at that time, $F_\mathrm{unspotted}$ is the estimated flux from the star showing a ``clean'' photospheric surface and $F_{\mathrm{norm},~i}$ is the resulting normalized flux at $t_i$. We modeled $F_{\mathrm{star},~i}$, the stellar flux local to the transit, by fitting a low-order (typically degree 2) polynomial to out-of-transit (OOT) data on either side of the transit. Accurate determination of $F_\mathrm{unspotted}$ was more challenging; it is difficult to determine when, if ever, we observe a clean disk. We took the maximum observed flux over each quarter (or month for short cadence photometry) as $F_\mathrm{unspotted}$ for that interval. To avoid an estimate biased by transient brightening events, the maximum was taken from a running average of the observed flux. This approach assumes that any variation in stellar brightness over the quarter is caused by starspots and not by an overall change in luminosity.

An example of the fit of a local stellar flux model, subsequent normalization of the transit, fit to a transit model, and resulting residuals is shown in Figure \ref{fig: lc normalization}.

\subsection{Changes in Crowding}
\label{sec: crowding}
Contamination of the photometric aperture by light from other stars also complicates accurate measurement of the transit depth. The PDC-MAP pipeline corrects this ``crowding,'' but time-varying errors in this procedure can produce changes in the measured transit depth. The process is analogous to star spots; the flux blocked by the planet is constant, but the estimated total flux from the star changes, yielding a change in the normalized transit depth. \citet{2013ApJ...774L..19V} found that the transit depth of the hot Jupiter HAT-P-7b varied by ${\sim}1\%$ from season to season,\footnote{The \textit{Kepler Space Telescope} performed an attitude adjustment four times per year to keep its solar panels pointed sunward. This causes each target to fall on a different CCD each ``season.''} orders of magnitude higher than the observed precision of transit depth measurements within a season. The authors suggested several possible causes, including incorrect crowding correction. Subsequently, \citet{2015A&A...576A..11G} found significant seasonal changes in transit depths measured from the PDC-MAP photometry of Kepler-423 (KOI 183). Since the changes in transit depths were highly correlated ($p = 0.15\%$) with the quarterly\footnote{Each change in \textit{Kepler} season marks a new quarter, so data from Q1 and Q5 are from the same season, one year apart.} crowding metric and were absent in the uncorrected SAP photometry, these authors attributed the changes entirely to the PDC-MAP crowding correction.

To screen for transit depth variations induced by crowding, we followed \citeauthor{2015A&A...576A..11G}, and checked for correlation between measured transit depths and the quarterly crowding metric. Additionally, for each candidate, we generated two normalized light curves, using the default PDC-MAP quarter-to-quarter crowding correction and a constant seasonal crowding value.

\subsection{Measuring Transit Depth}
Transit models were then fit to the normalized light curves in a two-step process. We first phase-folded ${\sim} 20$ transits and fit a typical quadratic limb darkening transit model \citep{2002ApJ...580L.171M} using $\chi^2$ minimization. The initial values were taken from the NASA Exoplanet Archive.\footnote{\url{http://exoplanetarchive.ipac.caltech.edu}} 
Then, to extract the transit depths, we fixed the limb darkening model and orbital parameters, and fit between one and a few transits with only the radius ratio, $p$, and transit mid-time, $t_0$, as free parameters.
The estimated $1 \sigma$ error in each depth measurement was determined from propagation of the PDC pipeline estimated photometric error. The estimated transit depth errors varied from star to star, with a typical value of ${\sim} 1\%$.
Transits were rejected if the data coverage over the transit was too sparse or large data gaps prevented an accurate fit to the out-of-transit flux.

\section{Determination of Oblateness}
\label{sec: findoblateness}
We examined the time series of measured transit depths for each candidate for signs of transit depth variations attributable to the spin precession of an oblate planet. To begin, we converted the series of measured transit depths into a time series of fractional transit depth variation, $\mathrm{T} \delta \mathrm{V}$, from the mean observed depth.\footnote{This definition is slightly different from that introduced in \citet{2010ApJ...716..850C}, who normalized by the minimum transit depth.}
Because transit depths may vary systematically from season to season in Kepler photometry \citep{2013ApJ...774L..19V}, we normalized the depths from each season by the mean of the observed depths from that season:
\begin{align}
\label{eq: tdeltav}
{\mathrm{T} \delta \mathrm{V}}_{s} = \frac{{\delta_\mathrm{obs,s}} - \overline{\delta_\mathrm{obs,s}}}{\overline{\delta_\mathrm{obs,s}}} \mathrm{,}
\end{align}
where the index $s$ indicates the season.

We constructed a corresponding model of the relative transit depth variation, {\tdv}$_m$, using Equation \eqref{eq: delta} in Section \ref{sec: oblateness-signal}, in which we substituted for $\phi$ and $\epsilon$ using \eqref{eq: phi} and \eqref{eq: ellipticity}, respectively.
As can be seen from these equations, this model has the precession period, $P_\mathrm{prec}$, oblateness, $f$, obliquity, $\theta$, transit inclination, $i$, and a phase offset, $t_0$ as free parameters. For a given choice of these parameters, we used Equation \eqref{eq: delta} to calculate the modeled transit depth, $\delta_m$, at the observed transit times. We then normalized the modeled depths from each season by the mean of the modeled depths from that season as in Equation \eqref{eq: tdeltav}.
Because we expect the planet-to-star radius ratio, $k = R_\mathrm{eq}/R_*$, and the stellar limb darkening model to remain constant over the observation period, this normalization by $\bar{\delta}$ makes both the observed and modeled $\mathrm{T} \delta \mathrm{V}$ series independent of these parameters.

\subsection{Fitting the $\mathrm{T} \delta \mathrm{V}$ Model}
Fits of this model to transit depth time series are generally not unique. High oblateness, for example, can be compensated by very low obliquity or very slow precession. We explored this parameter space using \texttt{emcee}, a Markov chain Monte Carlo Ensemble sampler written by \citet{emcee}. We used uniform priors over the appropriate physical range (e.g. $f \in [0, 1]$) for every model parameter but the transit inclination. The inclinations of our candidates were constrained by a normal prior based on the results of previous studies. For Kepler-39b we used $\mathcal{N}(89.23^{\circ},~0.014)$ \citep{2015AA...575A..85B}, while for Kepler-427b our prior was $\mathcal{N}(89.50^{\circ},~0.25)$ \citep{2014AA...572A..93H}.

We also restricted the precession period. First, we imposed a lower bound of 2 years. 
This is lower than expected for a Saturn-like planet on the inner-edge of our period range with modest rotational oblateness, $P_\mathrm{rot} = 18 \ \mathrm{hours} \Rightarrow f \approx 0.05$, and reduced the likelihood of overfitting to sparse data or detecting a spurious signal associated with \textit{Kepler's} ${\sim} 1$ year orbit. 
Second, because of the finite duration of the observations, a planet model with significant oblateness can match even unvarying transit depth measurements if compensated by a long enough precession period. 
To avoid this degeneracy, we required that the precession period be short enough that we could have observed a rise and fall in the measured transit depths. Given the ${\sim} 4$ year duration of the \textit{Kepler} mission, and that the transit depth variations peak twice per precession period, our final requirement was $P_\mathrm{prec} \ \mathrm{[years]} \in [2, 16]$.

To test the robustness and reliability of our MCMC approach, for each candidate we performed two injection and recovery tests. The first test was a spherical planet. Because oblateness is a positive-definite quantity, our marginalized posterior distributions of $f$ will have positive bias (see, for example, \citet{2011MNRAS.410.1895Z}). The spherical planet test gauged the extent of the bias and probed for other possible false positive signals. The second injected planet was Saturn-like. It had oblateness $f = 0.1$, obliquity $\theta = 30^\circ$, the same transit inclination as the real candidate, and a precession period determined by Equation \eqref{eq: precession period}, assuming the same orbital period as the candidate, a 10 hour rotation period and $\lambda / J_2 = 13.5$. In both cases, the planets were scaled so that the transit depth matched that measured for the real candidate.

We ran our MCMC analysis with 200 parallel chains taking 15,000 steps. We discarded the first ${\sim} 10000$ steps as the burn in period. 
We assessed convergence by inspection of the walker trajectories and by splitting the remaining part of the chain in half and comparing the resulting posteriors.
The resulting MCMC model was determined by taking the median of the marginalized posterior distribution of each parameter with the 16th and 84th percentiles taken as approximate $1 \sigma$ bounds.

\subsection{Statistical Significance}
The likelihood function was calculated using the errors estimated for each transit as described in Section \ref{sec: delta measurement}. These estimates, however, may understate the true error. We performed a series of injection and recovery tests to assess our measurement accuracy and gauge the resulting false positive rate.

\subsubsection{Simulated Transit Light Curves}
When \citet{2010ApJ...716..850C} investigated the detectability of transit depth variations in simulated \textit{Kepler} photometry, they used a Gaussian white noise model with a simulated transit light curve. They found that a single transit's depth could be recovered to within $\approx 0.6\%$ for a star with a \textit{Kepler} magnitude of 13. With the full 4 years of data in hand, we attempted to improve this estimate by injecting transits into real \textit{Kepler} photometry.

For each candidate, we constructed a series of simulated \textit{Kepler} data sets using the photometry for that star. First, we removed all known transits from the data. We considered the remaining photometric data to consist of the true stellar flux plus some noise-induced offset: $F_\mathrm{meas} = F_\mathrm{true} + F_\mathrm{offset}$. 
We calculated the running average of the photometry and took this as our model for the stellar flux, $F_\mathrm{true}$. We then added artificial \citet{2002ApJ...580L.171M} transit models to the stellar flux: $F_\mathrm{true} \rightarrow F'_\mathrm{true}$. The simulated data was then generated by applying the same offset that was present in the real data: $F_\mathrm{sim} = F'_\mathrm{true} + F_\mathrm{offset}$. 

After generating the artificial \textit{Kepler} observations, we then fed the data into our analysis pipeline. We tested our analysis using both a constant and variable planet-to-star radius ratio to generate the injected transit models. For each star we generated ${\sim} 1000$ simulated transits and compared our pipeline's estimated error, derived from the \textit{Kepler} pipeline's reported photometric uncertainty, to the actual accuracy obtained from comparing the recovered transit parameters to the injected ones. We found that propagation of the photometric precision tends to underestimate the transit depth measurement error. An example of the error distribution is shown in Figure \ref{fig: KOI423 SimErrors} for Kepler-39 (KOI 423).

\subsubsection{Simulated Transit Depth Series}
Next, we investigated the false positive rate for oblateness detection using synthetic $\mathrm{T} \delta \mathrm{V}$ data. We generated $\mathrm{T} \delta \mathrm{V}$ time series for a spherical planet in two ways, by sampling from the error distributions described above and by bootstrapping from the observed \tdv series. To account for the possibility of seasonally correlated data, we used block bootstrapping; synthetic data for each season was resampled from that season's observations. For each time series, we then calculated the $\Delta\chi^2 = \chi^2_\mathrm{sph} - \chi^2_\mathrm{obl}$ value for the minimum-$\chi^2$ oblate planet model and a spherical planet model. These values were used to create both a simulated and bootstrapped distribution of $\Delta \chi^2$ values for that star.
Given an oblate model fit to \textit{Kepler} data and two $\Delta \chi^2$ distributions from simulated \tdv ~series, we calculate p-values from the fraction of simulated $\Delta \chi^2$ values higher than the one obtained by the model and report the more conservative estimate.

\section{Results}
\label{sec: results}
Of the 9 candidates matching our period criteria and having $R_p > 6 \ R_\oplus$, we selected two, Kepler-39b (KOI 423.01) and Kepler-427b (KOI 192.01) for close evaluation based on their large transit depths (${\sim}1\%$) and independently measured masses. The results are presented below and summarized in Table \ref{table: MCFit}.

\subsection{Kepler-39b (KOI 423.01)}
\citet{2014ApJ...796...67Z} conducted a search for oblate planets using \textit{Kepler} short cadence photometry to identify deviations from the light curve of a perfectly spherical transiting planet. The ${\sim} 20 \ M_J$ Kepler-39b (KOI 423.01) was the only object found with likely non-zero oblateness. Though the measured projected oblateness was high, $f_\perp = 0.22 \pm 0.11$, the authors cautioned that the finding might not be robust due to inconsistency in the best fit models when different subsets of short-cadence transits were examined.

\begin{figure}[t]
	\centering
	\includegraphics[width=0.5\textwidth]{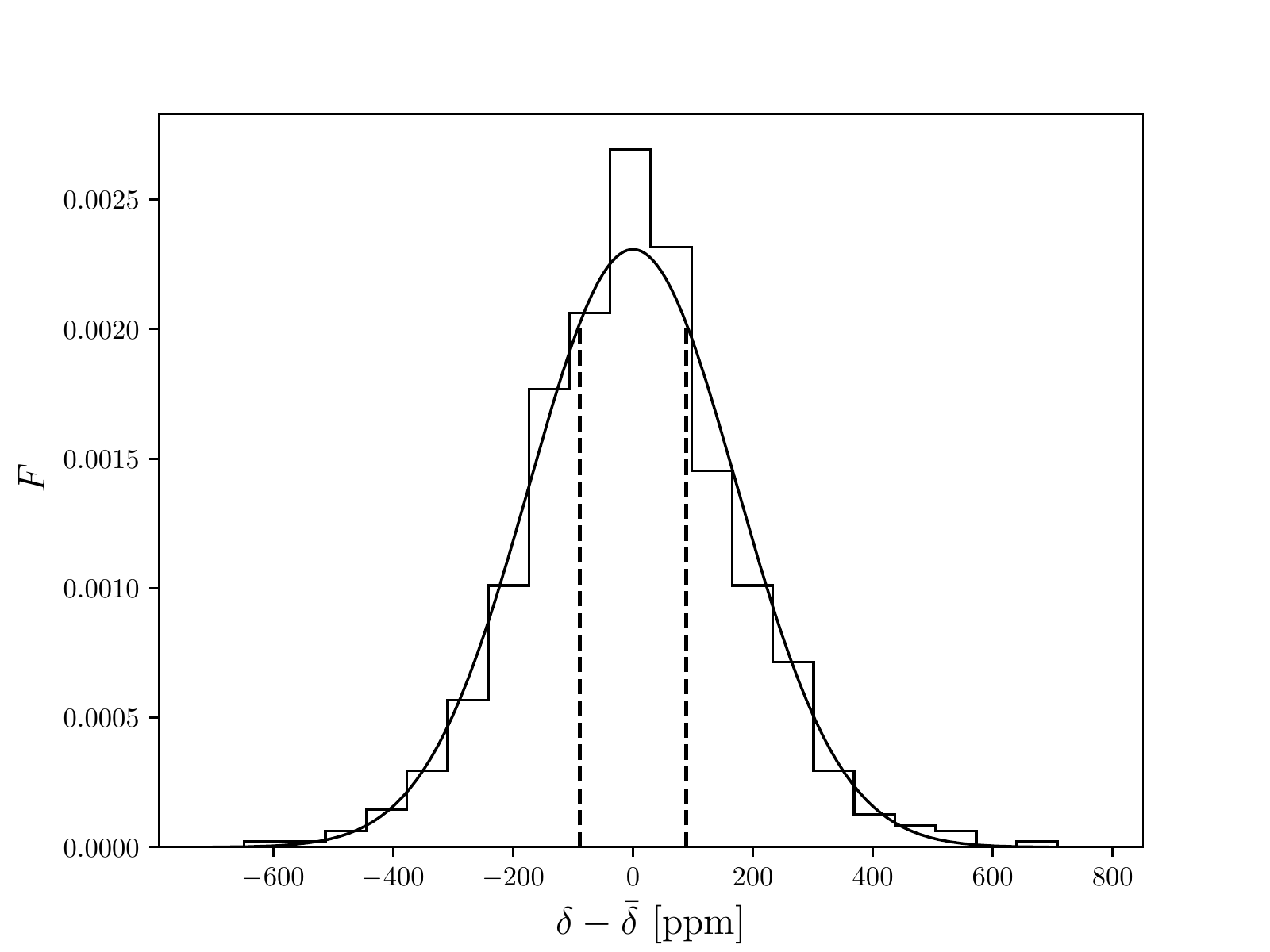}
	\caption{\small Normalized histogram of errors in depth measurements from 700 transits injected into the \textit{Kepler} photometry of Kepler-39. The injected planet was spherical with the same projected area as Kepler-39b. The solid black curve is a Gaussian fit to the distribution with $\sigma = 172.80~\mathrm{ppm}$ and $\mu \approx 10^{-13}$. The dashed lines are at ${\sim} 1 \%$ of the transit depth of Kepler-39b ($\pm 89 \ \mathrm{ppm}$)\textemdash the approximate magnitude of an oblate signal. The 68th and 95th-percentile estimated depth errors returned by our analysis pipeline on the same transits were $122.88 \ \mathrm{ppm}$ and $162.83 \ \mathrm{ppm}$, respectively.}
	\label{fig: KOI423 SimErrors}
\end{figure}

With this tentative detection, and an orbital period of approximately 21 days, Kepler-39b presented a promising target for the $\mathrm{T} \delta \mathrm{V}$ method. We fit normalized transit models to 46 of the 69 transits spanning 3.93 years of long-cadence \textit{Kepler} photometry. Kepler-39 is a $1.29 \ M_{\odot}$ star with a Kepler band magnitude of $14.33$ \citep{2015AA...575A..85B}. Due to the relative faintness of the star, the long cadence data has a photometric precision of ${\sim} 200~\mathrm{ppm}$. The standard deviation of errors in our $\mathrm{T} \delta \mathrm{V}$ signal, as determined by our injection and recovery test, was $\sigma = 173~\mathrm{ppm}$ (see Figure \ref{fig: KOI423 SimErrors}). We found no correlation between the measured transit depth and the \textit{Kepler} quarterly crowding metric.

\begin{figure}[t]
	\centering
	\includegraphics[width=0.5\textwidth]{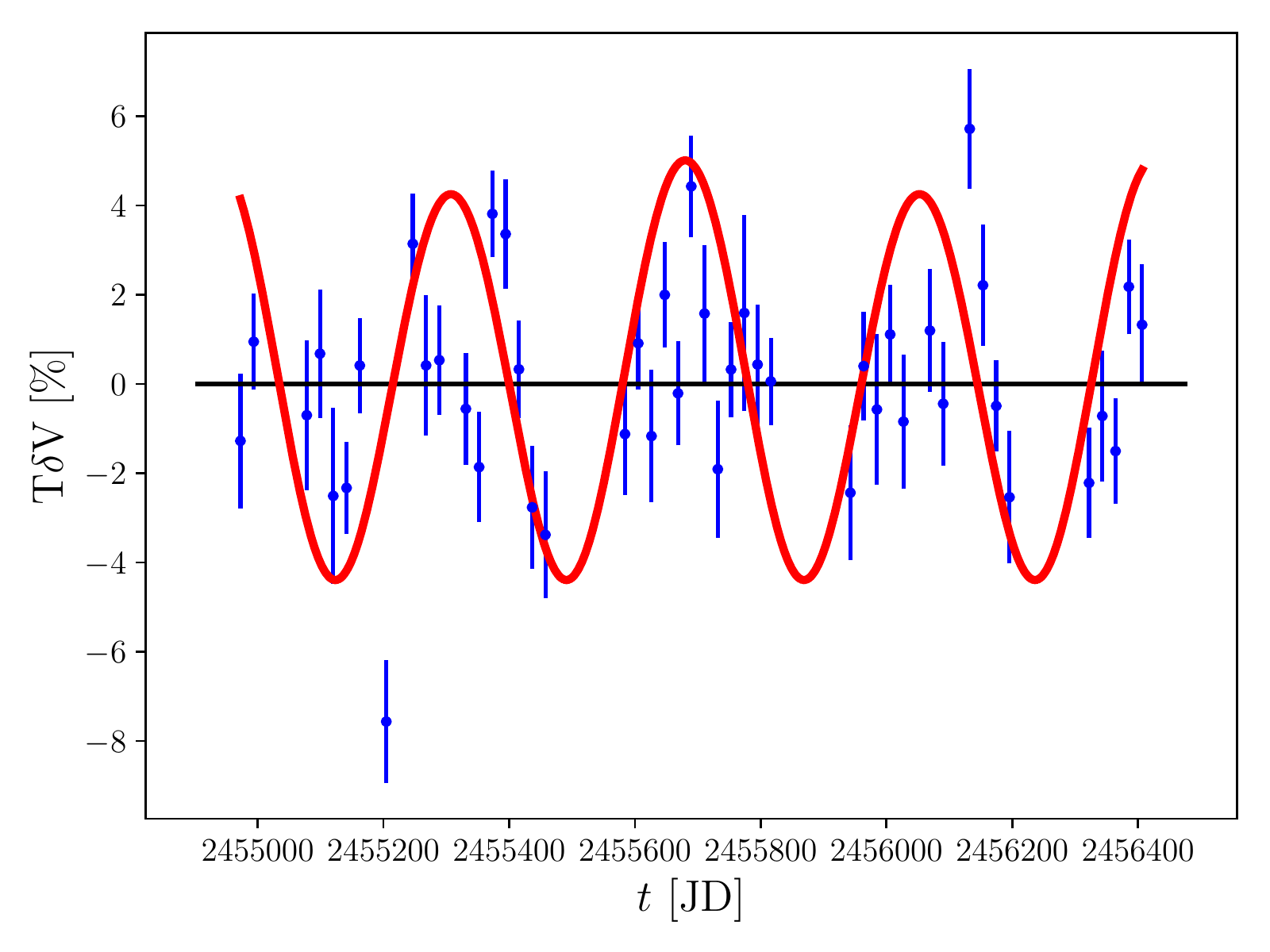}
	\caption{\small Transit depth variation relative to mean of all observed depths for Kepler-39b. The data are plotted in blue with estimated $1\sigma$ uncertainty. The red solid line is the best-fit oblate planet model.}
	\label{fig: KOI423 Depths}
\end{figure}

The measured relative $\mathrm{T} \delta \mathrm{V}$ and the best-fit model are plotted in Figure \ref{fig: KOI423 Depths}. For visual clarity, we omitted the separate seasonal normalizations and show the variations relative to the mean of all the observed depths. The MCMC analysis yielded modest oblateness, $f = 0.13^{+0.12}_{-0.04}$, with a rapid precession period of $P_\mathrm{prec} = 2.04^{+0.02}_{-0.02} \ \mathrm{years}$. The full model is detailed in Table \ref{table: MCFit} and the posterior distributions are shown in Figure \ref{fig: KOI423 Data Corner}. The precession period is tightly bound at the lower edge of the range of allowed periods. The joint $f$-$\theta$ distribution shows the expected degenerate relationship (Equation \ref{eq: f theta degeneracy}), with zero oblateness excluded and low oblateness allowed only at very high obliquities.

\begin{figure}[t]
	\centering
	\includegraphics[width=0.5\textwidth]{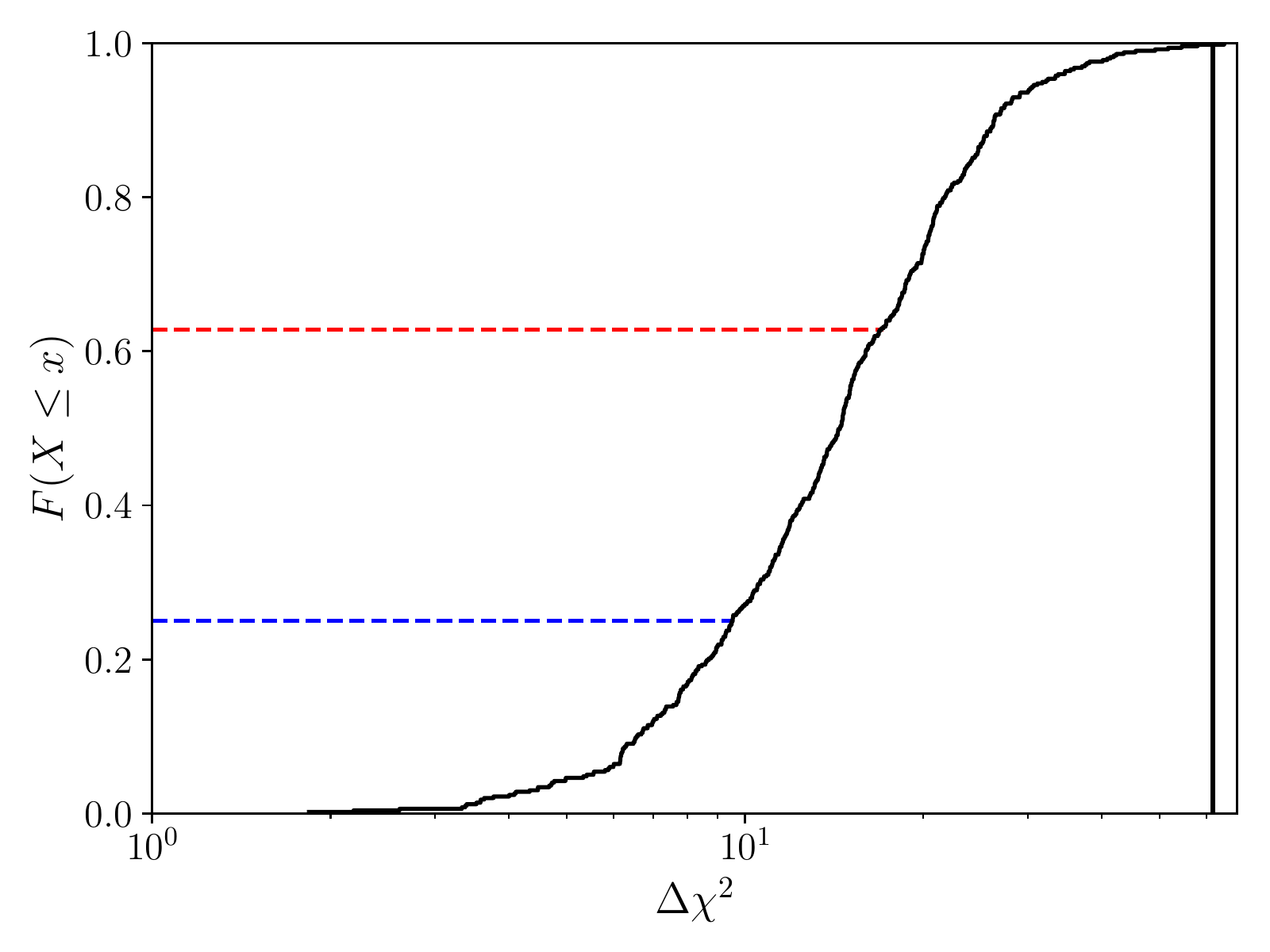}
	\caption{\small Cumulative distribution of $\Delta \chi^2$ values for oblate planet model fits to simulated $\mathrm{T} \delta \mathrm{V}$ measurements of a spherical planet around Kepler-39. The black vertical bar is the score obtained by the fit to the real \textit{Kepler} data ($\Delta \chi^2 = 61.54$), corresponding to a p-value of $p = 1 - F = 0.002$. The horizontal dashed lines show the p-values obtained for the injection and recovery of the simulated spherical planet (red; $p = 0.37$), and the simulated Saturn-like planet (blue; $p = 0.75$).}
	\label{fig: KOI423 DeltaChi2 CDF}
\end{figure}

The cumulative distribution of $\Delta \chi^2$ values for the simulated ensemble of spherical planet models is shown in Figure \ref{fig: KOI423 DeltaChi2 CDF}. The recovered oblate model scored $61.54$, corresponding to $p = 0.002$, or significant at the $3.09 \sigma$ level.

\begin{figure*}[h]
	\centering
	\includegraphics[width=1.0\textwidth]{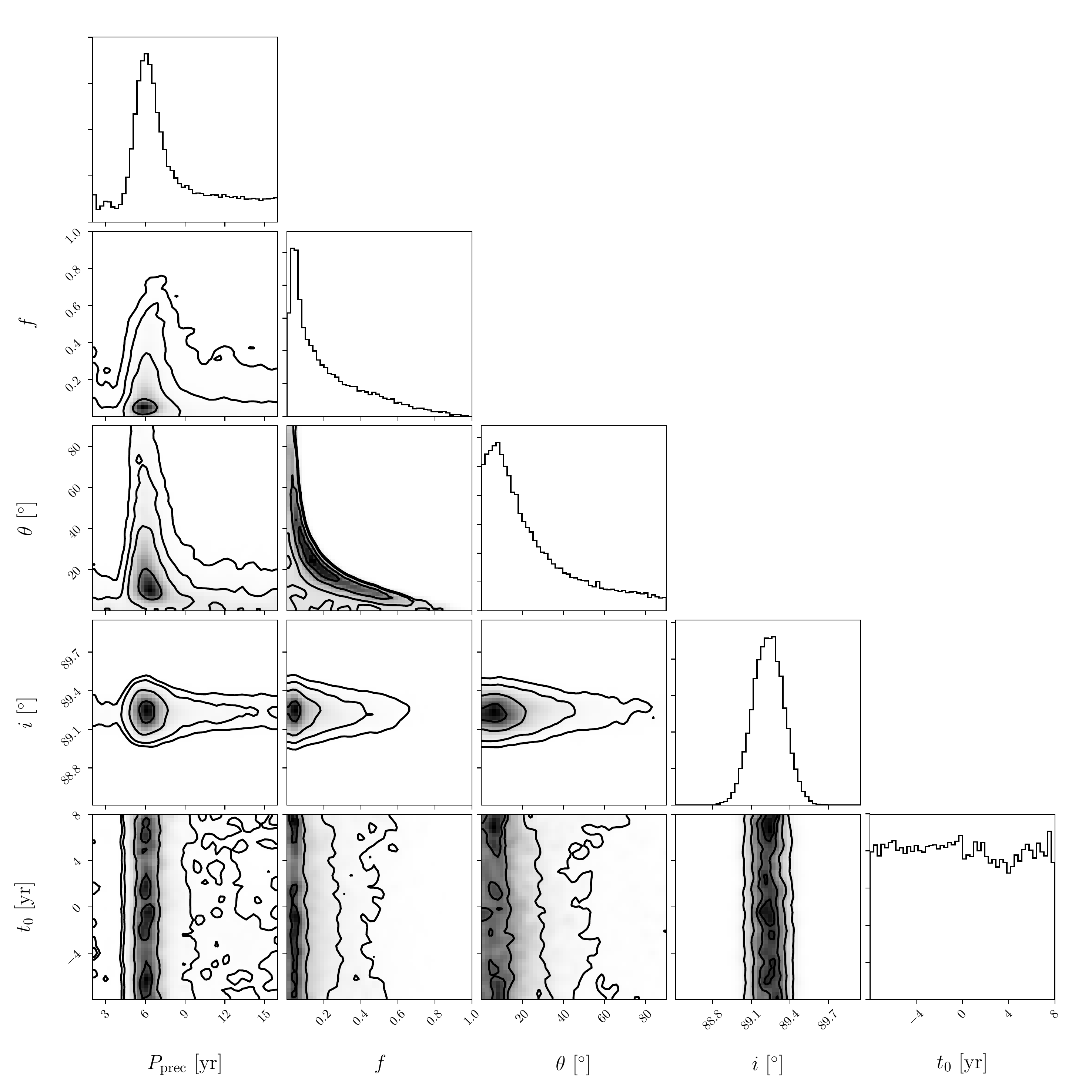}
	\caption{\small Posterior probability distributions from MCMC analysis of transit depths from spherical planet transits injected into Kepler-39 photometry. Contours are drawn at $0.5 \sigma$ intervals from $0.5 - 2.0 \sigma$ and the plots have been smoothed to remove noisy features.}
	\label{fig: KOI423 Null Corner}
\end{figure*}

We created similar $\Delta \chi^2$ distributions based on the simulated data we generated for injected spherical and Saturn-like planets. The p-values obtained from these distributions are also shown in Figure \ref{fig: KOI423 DeltaChi2 CDF}. The recovered model from the injected spherical planet had $p = 0.37$, or $0.90\sigma$, indicating an oblate model provides no statistical advantage over a spherical planet model. The absence of a robust detection is corroborated by the posterior distributions shown in Figure \ref{fig: KOI423 Null Corner}.
For $i$ and $t_0$, we recovered our Gaussian and uniform priors, respectively. The posteriors for $f$ and $\theta$ display the expected degeneracy while the joint distribution shows that the data support models with near zero oblateness even for modest obliquity. The $P_\mathrm{prec}$ posterior spans the allowed range but has a broad peak in probability density centered near 6 years.

The recovered model for the Saturn-like planet was also not statistically significant, with $p = 0.75$, corresponding to $0.32 \sigma$. As before, the posterior distributions (Figure \ref{fig: KOI423 Null Corner}) reflect the absence of a detection, while in this case the $P_\mathrm{prec}$ posterior shows unexpected narrow peaks at periods of 2 and 3 years. The failure to detect the injected Saturn-like planet was not surprising, given that the injected period, $P_\mathrm{prec} = 30.35~\mathrm{years}$, exceeds the upper limit we imposed on the precession period, but the high probability density at low periods was unexpected. This anomaly in the precession period posterior for the Saturn-like case occurs at roughly the same period as the signal in the real photometry, suggesting that detection may be spurious.

\subsection{Kepler-427b (KOI 192.01)}
\begin{figure}[h]
	\centering
	\includegraphics[width=0.5\textwidth]{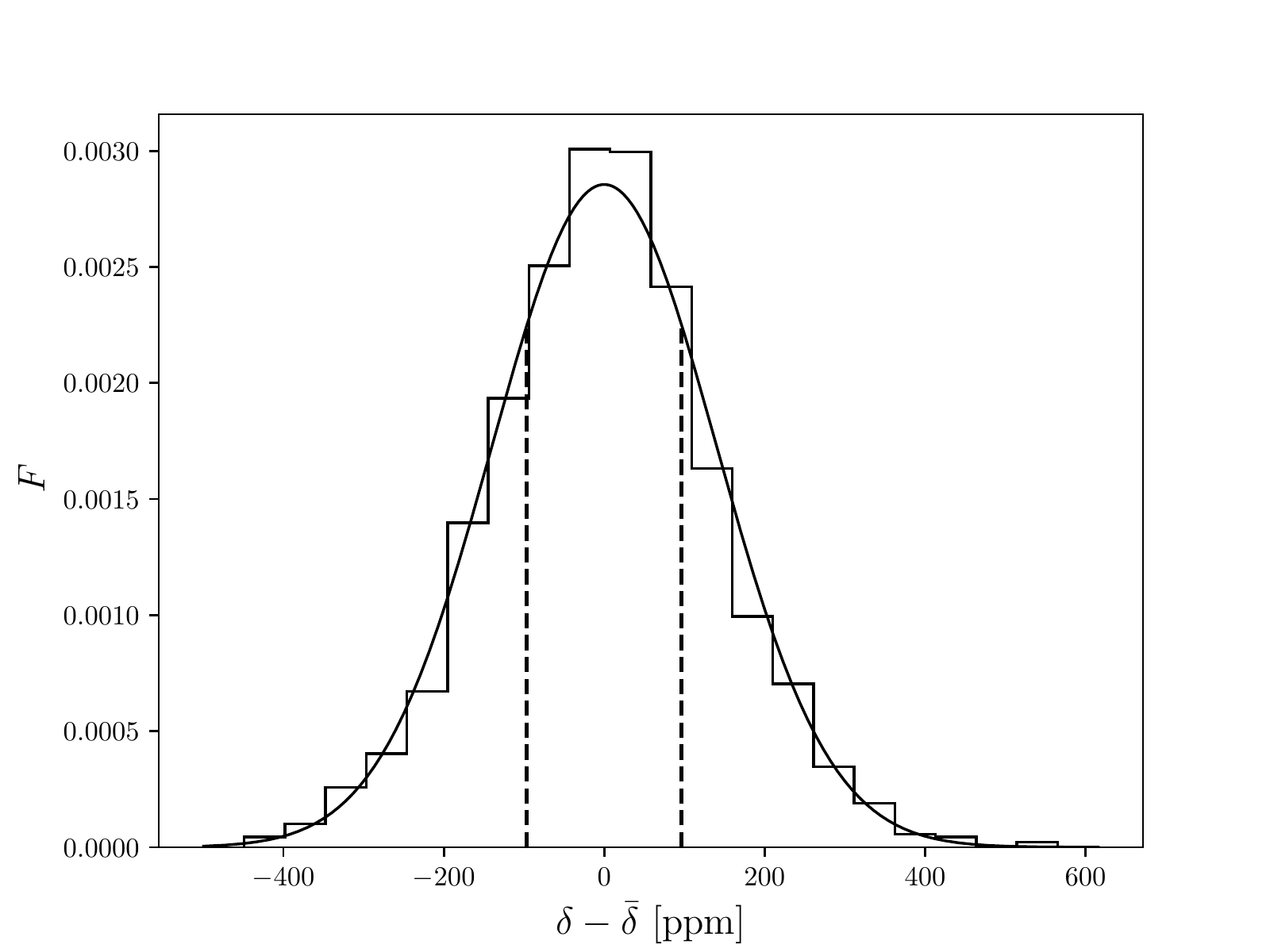}
	\caption{\small Normalized histogram of errors in depth measurements from 1764 transits injected into the \textit{Kepler} photometry of Kepler-427. The injected planet was spherical with the same projected area as Kepler-427b. The solid black curve is a Gaussian fit to the distribution with $\sigma = 139.73~\mathrm{ppm}$ and $\mu \approx -10^{-13}$. The dashed lines are at ${\sim} 1 \%$ of the transit depth of Kepler-427b ($\pm 97 \ \mathrm{ppm}$)\textemdash the approximate magnitude of an oblate signal. The 68th and 95th-percentile estimated depth errors returned by our analysis pipeline on the same transits were $82.20~\mathrm{ppm}$ and $104.74~\mathrm{ppm}$, respectively.}
	\label{fig: KOI192 SimErrors}
\end{figure}

Kepler-427 (KOI 192), is a $0.96 \pm 0.06 \ M_\odot$ star hosting a $0.29 \pm 0.09 \ M_J$ planet on a 10.3 day orbit \citep{2014AA...572A..93H}. Like Kepler-39, Kepler-427 is a relatively dim star with a Kepler band magnitude of $14.22$ \citep{2014AA...572A..93H}, resulting in photometric precision of approximately $185~\mathrm{ppm}$. From recovery of simulated transits we obtained a standard deviation of errors $\sigma = 140~\mathrm{ppm}$ (see Figure \ref{fig: KOI192 SimErrors}). Again we find no significant correlation between the \textit{Kepler} quarterly crowding metric and the observed transit depths.

\begin{figure}[h]
	\centering
	\includegraphics[width=0.5\textwidth]{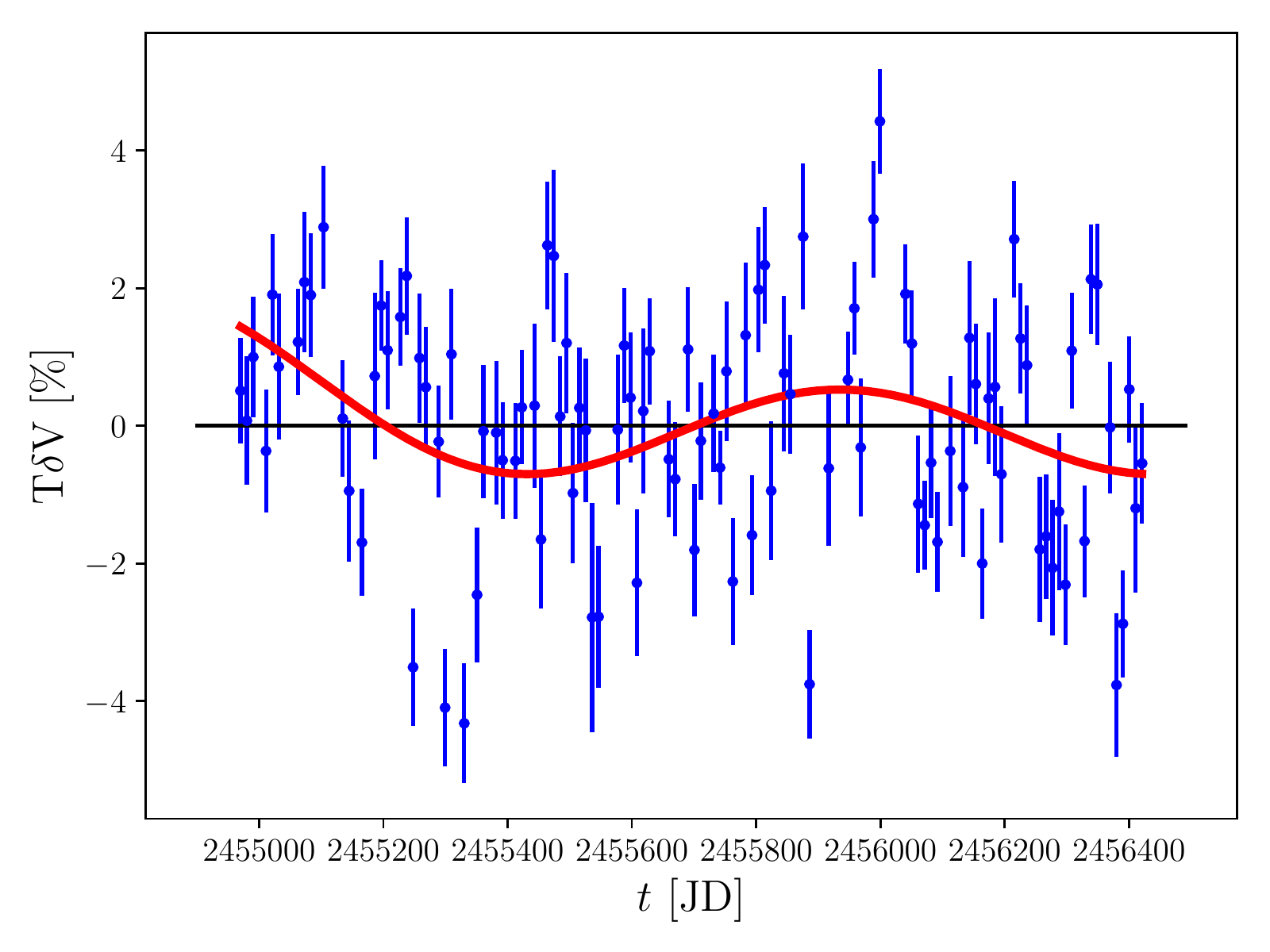}
	\caption{\small Transit depth variation relative to mean of all observed depths for Kepler-427b. The data are plotted in blue with $1\sigma$ estimated errors. The red solid line is the best-fit oblate planet model.}
	\label{fig: KOI192 Depths}
\end{figure}

\begin{figure}[h]
	\centering
	\includegraphics[width=0.5\textwidth]{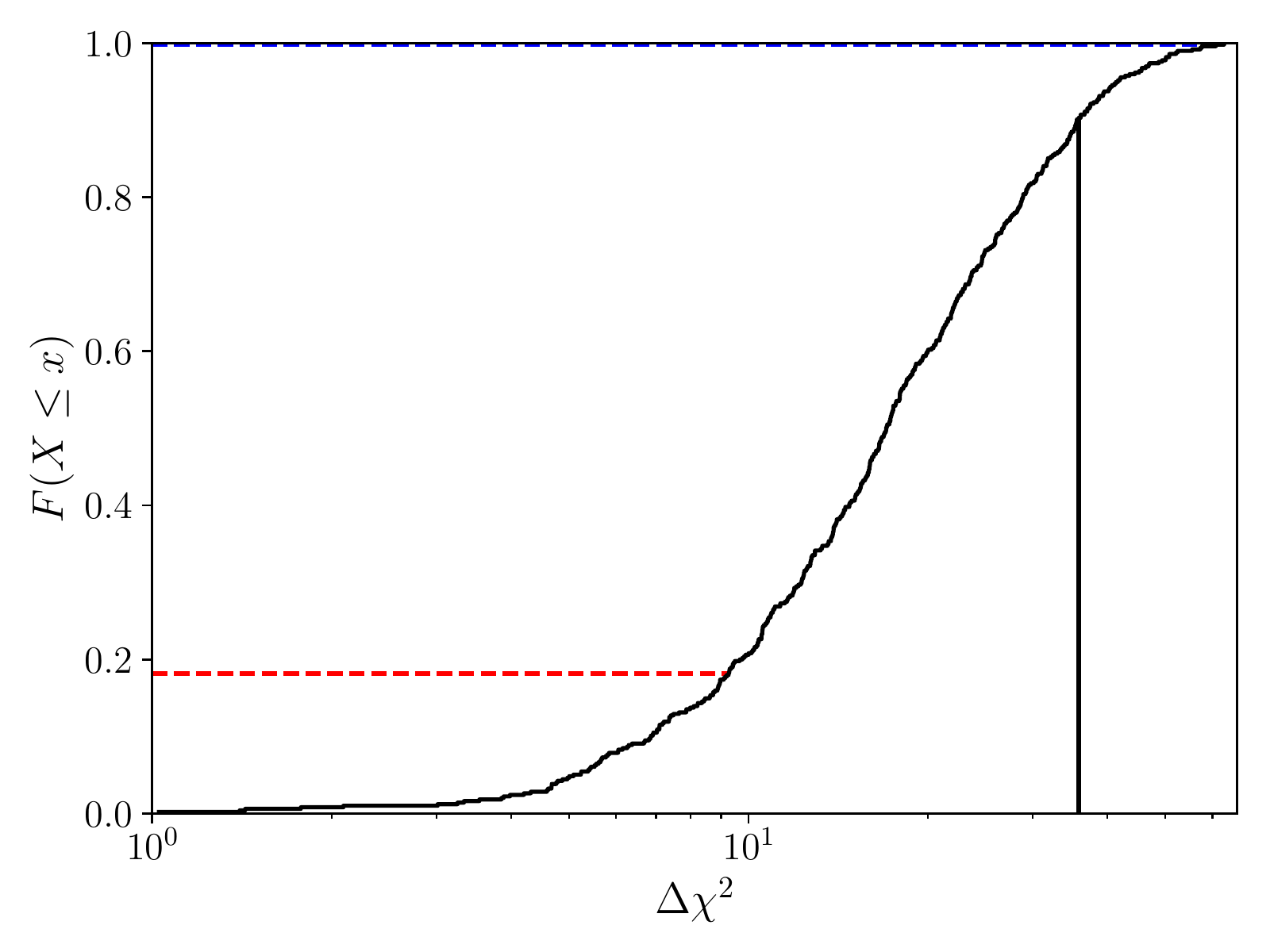}
	\caption{\small Cumulative distribution of $\Delta \chi^2$ values for oblate planet model fits to simulated $\mathrm{T} \delta \mathrm{V}$ measurements of a spherical planet around Kepler-427. The black vertical bar is the score obtained by the fit to the real \textit{Kepler} data ($\Delta \chi^2 = 35.80$), corresponding to a p-value of $p = 1 - F = 0.099$. The horizontal dashed lines show the p-values obtained for the injection and recovery of the simulated spherical planet (red; $p = 0.818$), and the simulated Saturn-like planet (blue; $p < 0.002$).}
	\label{fig: KOI192 DeltaChi2 CDF}
\end{figure}

\begin{figure*}[h]
	\centering
	\includegraphics[width=1.0\textwidth]{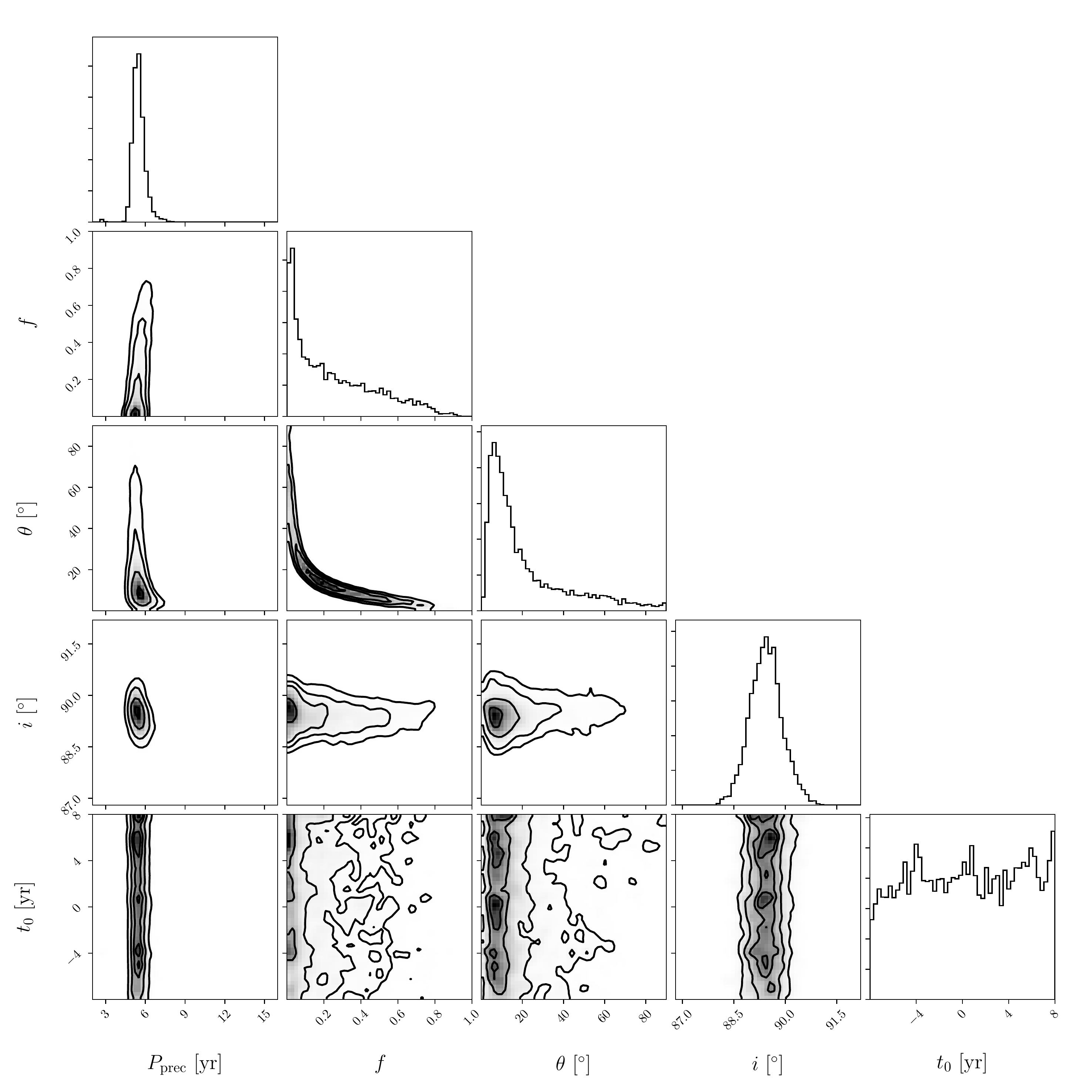}
	\caption{\small Posterior probability distributions from MCMC analysis of Kepler-427b transit depths. Contours are drawn at $0.5 \sigma$ intervals from $0.5 - 2.0 \sigma$ and the plots have been smoothed to remove noisy features.}
	\label{fig: KOI192 Data Corner}
\end{figure*}

We fit 104 long-cadence transits spanning nearly 4 years of \textit{Kepler} photometry, with the resulting relative $\mathrm{T} \delta \mathrm{V}$ measurements and the best-fit model plotted in Figure \ref{fig: KOI192 Depths}. 
The oblate planet model had $\Delta \chi^2 = 35.80$, which, in comparison to the distribution shown in Figure \ref{fig: KOI192 DeltaChi2 CDF}, yields $p = 0.099$, or $1.65 \sigma$. From the MCMC posterior distributions (Figure \ref{fig: KOI192 Data Corner}) we obtained $f =  0.19^{+0.32}_{-0.16}$ with a precession period of $P_\mathrm{prec} =  5.45^{+0.46}_{-0.37}~\mathrm{years}$.

\begin{figure*}[h]
	\centering
	\includegraphics[width=1.0\textwidth]{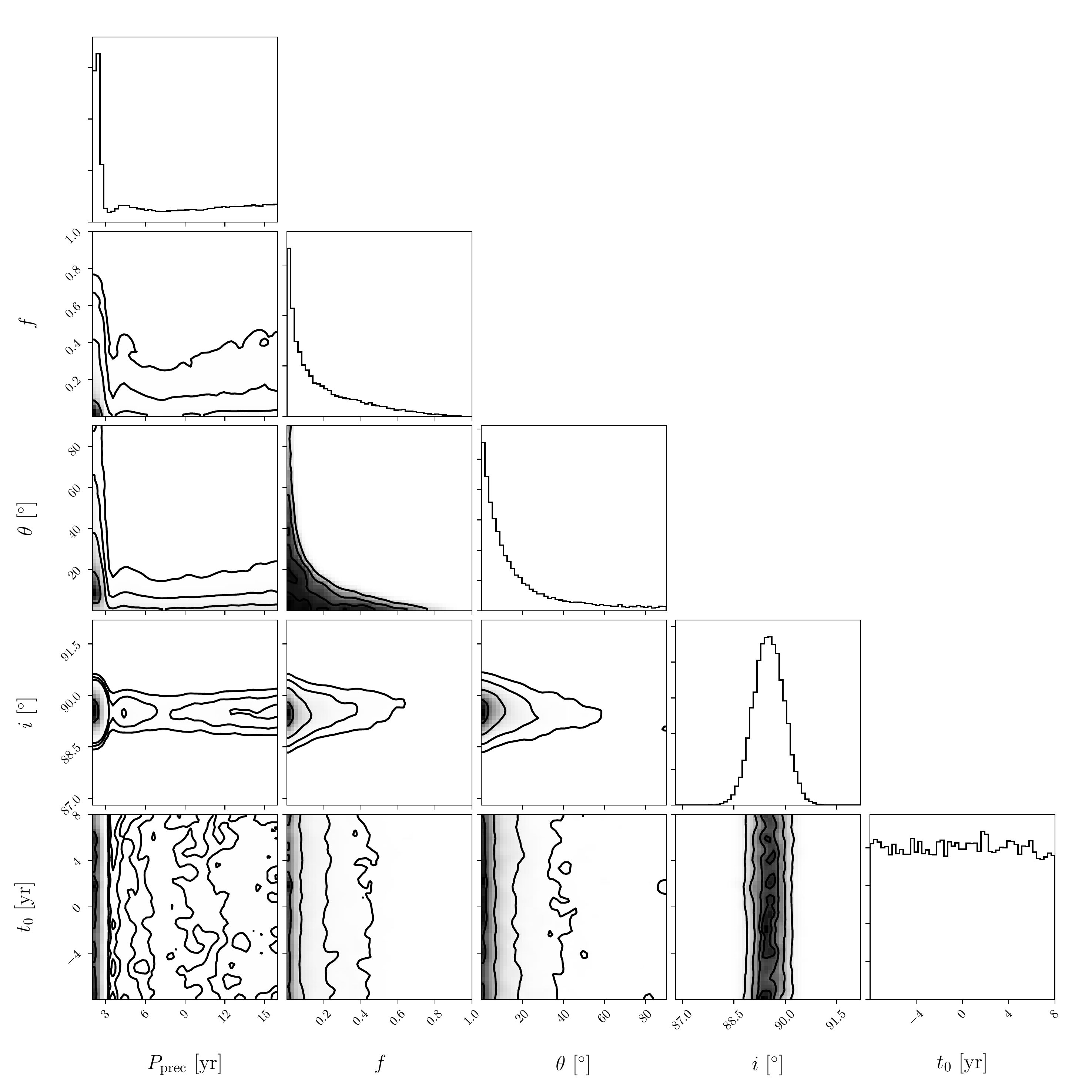}
	\caption{\small Posterior probability distributions from MCMC analysis of transit depths from spherical planet transits injected into Kepler-427 photometry. Contours are drawn at $0.5 \sigma$ intervals from $0.5 - 2.0 \sigma$ and the plots have been smoothed to remove noisy features.}
	\label{fig: KOI192 Null Corner}
\end{figure*}

The model recovered for the simulated spherical planet matches the expectation for unvarying transit depths. The posterior distributions (Figure \ref{fig: KOI192 Null Corner}) largely recover the input priors, with the familiar degenerate structure for $f$ and $\theta$. The posterior distribution of precession period shows a short-period peak analogous to that recorded in the Saturn-like simulation for Kepler-39b. 
The oblate model fit is not significant, scoring $p = 0.818$, or $0.23 \sigma$, indicating no improvement on a spherical model.

\begin{figure*}[h]
	\centering
	\includegraphics[width=1.0\textwidth]{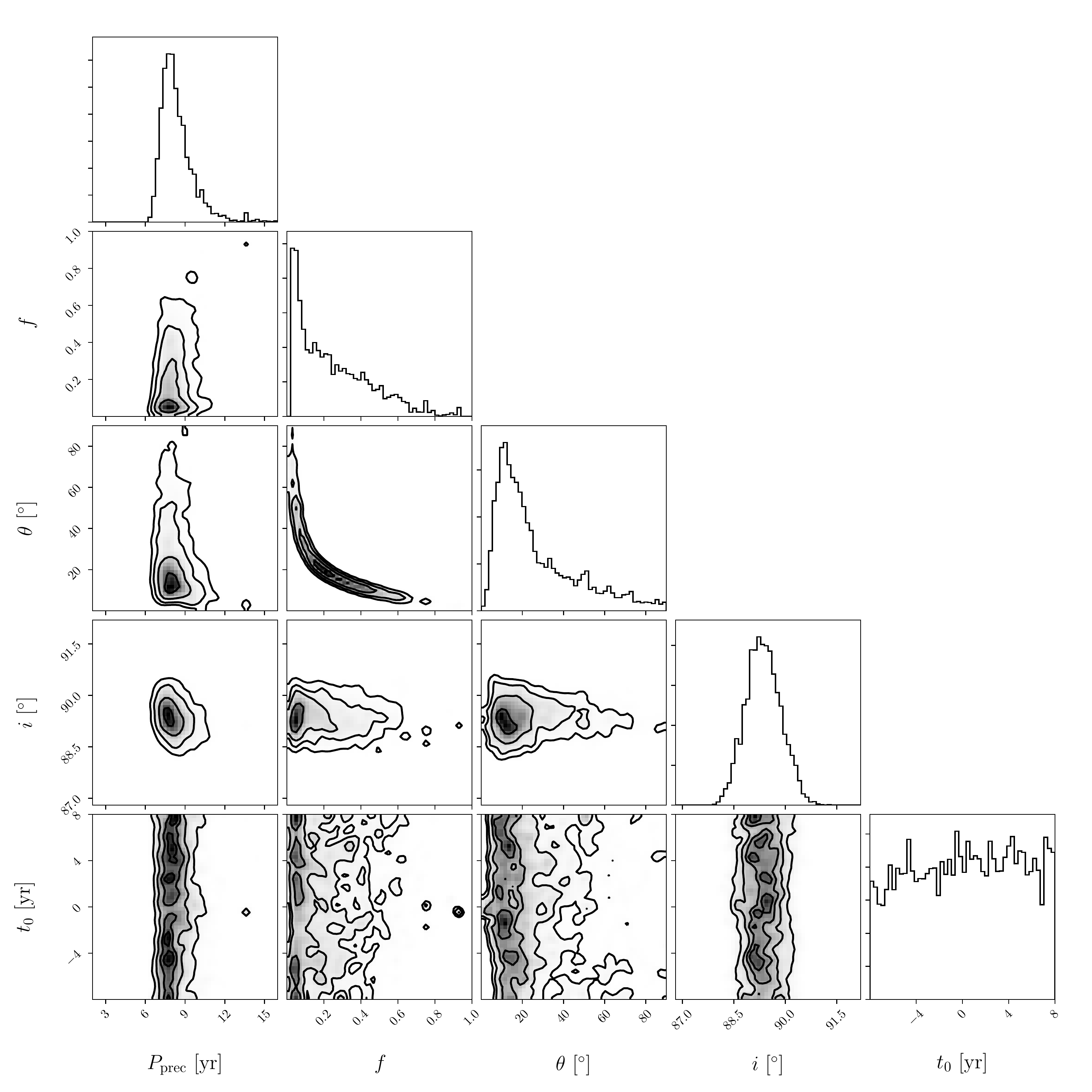}
	\caption{\small Posterior probability distributions from MCMC analysis of transit depths from Saturn-like planet transits injected into Kepler-427 photometry. Contours are drawn at $0.5 \sigma$ intervals from $0.5 - 2.0 \sigma$ and the plots have been smoothed to remove noisy features.}
	\label{fig: KOI192 Synth Corner}
\end{figure*}

In the case of the simulated Saturn-like planet, the oblate model is a highly significant improvement on a spherical planet fit, achieving $p < 0.002$ or ${>}3.09 \sigma$. 
In addition, with the exception of the precession period, all the input parameters were recovered in the estimated 68\% confidence interval (see Figure \ref{fig: KOI192 Synth Corner}). The precession period was only narrowly outside the interval. The poorly resolved period is likely due to the 4 year duration of $\mathrm{T} \delta \mathrm{V}$ data compared to the 7.23 year injected period.
We recovered $f \sin^2{\theta}$ to within 12.5\% of the input value, but, in part due to the $f$-$\theta$ degeneracy (Equation \ref{eq: f theta degeneracy}), we only recovered $f$ to within 50\% and $\theta$ to within 35\%.

\floattable
\begin{deluxetable}{ccccccc}
\tablewidth{\textwidth}
\tablecolumns{6}
\tabletypesize{\footnotesize}
\tablecaption{MCMC fits to $\mathrm{T} \delta \mathrm{V}$ measurements \label{table: MCFit}}
\tablehead{
\colhead{Planet} & \colhead{$P_\mathrm{prec} \ \mathrm{[year]}$} & \colhead{$f$} & \colhead{$\theta \ \mathrm{[^{\circ}]}$} & \colhead{$i \ \mathrm{[^{\circ}]}$} & \colhead{$t_0 \ \mathrm{[year]}$}
}
\startdata
Kepler-39b & $2.04^{+0.02}_{-0.02}$ & $0.13^{+0.12}_{-0.04}$ & $50.65^{+25.86}_{-21.03}$ & $89.24^{+0.12}_{-0.12}$ & $-0.12^{+5.44}_{-5.38}$ \\
\\
Kepler-427b & $5.45^{+0.46}_{-0.37}$ & $0.19^{+0.32}_{-0.16}$ & $13.62^{+29.11}_{-7.95}$ & $89.44^{+0.46}_{-0.46}$ & $0.31^{+5.42}_{-5.35}$ \\
\cutinhead{Kepler-39 (KOI 423.01) Simulated Planets}
Spherical (Injected) & \nodata & $ 0.0 $ & $ 0 $ & $ 89.23 $ & \nodata \\
\\
Spherical (Recovered) & $6.74^{+4.95}_{-1.44}$ & $0.15^{+0.30}_{-0.11}$ & $17.60^{+31.14}_{-12.21}$ & $89.23^{+0.12}_{-0.12}$ & $-0.21^{+5.66}_{-5.30}$ \\
\\
Saturn-like (Injected) & $ 30.35 $ & $ 0.10 $ & $ 30.0 $ & $ 89.23 $ & $ 13.71 $ \\
\\
Saturn-like (Recovered) & $7.46^{+6.08}_{-4.59}$ & $0.14^{+0.30}_{-0.11}$ & $11.69^{+27.60}_{-8.85}$ & $89.23^{+0.12}_{-0.12}$ & $0.02^{+5.44}_{-5.41}$ \\
\cutinhead{Kepler-427 (KOI 192.01) Simulated Planets}
Spherical (Injected) & \nodata & $ 0.0 $ & $ 0 $ & $ 89.50 $ & \nodata \\
\\
Spherical (Recovered) & $5.60^{+7.65}_{-3.30}$ & $0.12^{+0.28}_{-0.10}$ & $9.64^{+24.61}_{-7.53}$ & $89.52^{+0.44}_{-0.44}$ & $0.00^{+5.36}_{-5.46}$ \\
\\
Saturn-like (Injected) & $ 7.23 $ & $ 0.10 $ & $ 30.0 $ & $ 89.50 $ & $ -1.26 $ \\
\\
Saturn-like (Recovered) & $8.11^{+1.30}_{-0.78}$ & $0.20^{+0.26}_{-0.15}$ & $19.68^{+28.82}_{-10.07}$ & $89.35^{+0.53}_{-0.50}$ & $0.43^{+5.05}_{-5.48}$ \\
\enddata
\tablecomments{The reported values are the medians along with the 16th and 84th percentiles.}
\end{deluxetable}

The recovery of an injected candidate and the absence of a false positive detection in the null simulation indicate that the $1.65 \sigma$ confidence level of the recovered signal in the real \textit{Kepler} data is likely valid.

To confirm that the observed change in transit depth was not a byproduct of our transit normalization technique, we repeated our analysis with the starspot correction (Section \ref{sec: norm with spots}) removed. Kepler-427 only showed starspot-like brightness fluctuations in a few quarters. As a consequence, the average change in recovered transit depths was only ${\sim} 8~\mathrm{ppm}$, less than 0.1\% of the transit depth. The change in measured depths affected the model recovered from our MCMC analysis, but for each parameter, the recovered values with or without the starspot correction are mutually consistent\textemdash the median value obtained with one method is within the estimated $1\sigma$ uncertainty of the median value obtained with the other normalization. In particular, the measured precession period changed by ${\sim}$0.5\%, while $f \sin^2{\theta}$ changed by ${\sim}$7\%. Finally, removing the starspot correction marginally diminishes the statistical significance of the detection, yielding $p = 0.112$ or $1.59\sigma$. Based on these findings, we conclude that the observed $T \delta V$ signal is not an artifact of our normalization procedure.

\section{Discussion}
\label{sec: discussion}
In this paper, we present the first attempt to detect the rotational oblateness of an exoplanet through long-term changes in transit depth. We examined confirmed \textit{Kepler} planets with periods between 10 and 30 days and radii greater than $6~R_{\oplus}$ for signs of transit depth variations consistent with the spin-axis precession of an oblate planet. 
Of the 9 planets matching our criteria, we selected two for close analysis, Kepler-39b (KOI 423.01) and Kepler-427b (KOI 192.01). We searched for an oblate signature using a Markov Chain Monte Carlo approach and assessed the significance of the recovered oblate planet fits.

\subsection{Kepler-39b (KOI 423.01)}
\begin{figure*}[h]
	\centering
	\includegraphics[width=1.0\textwidth]{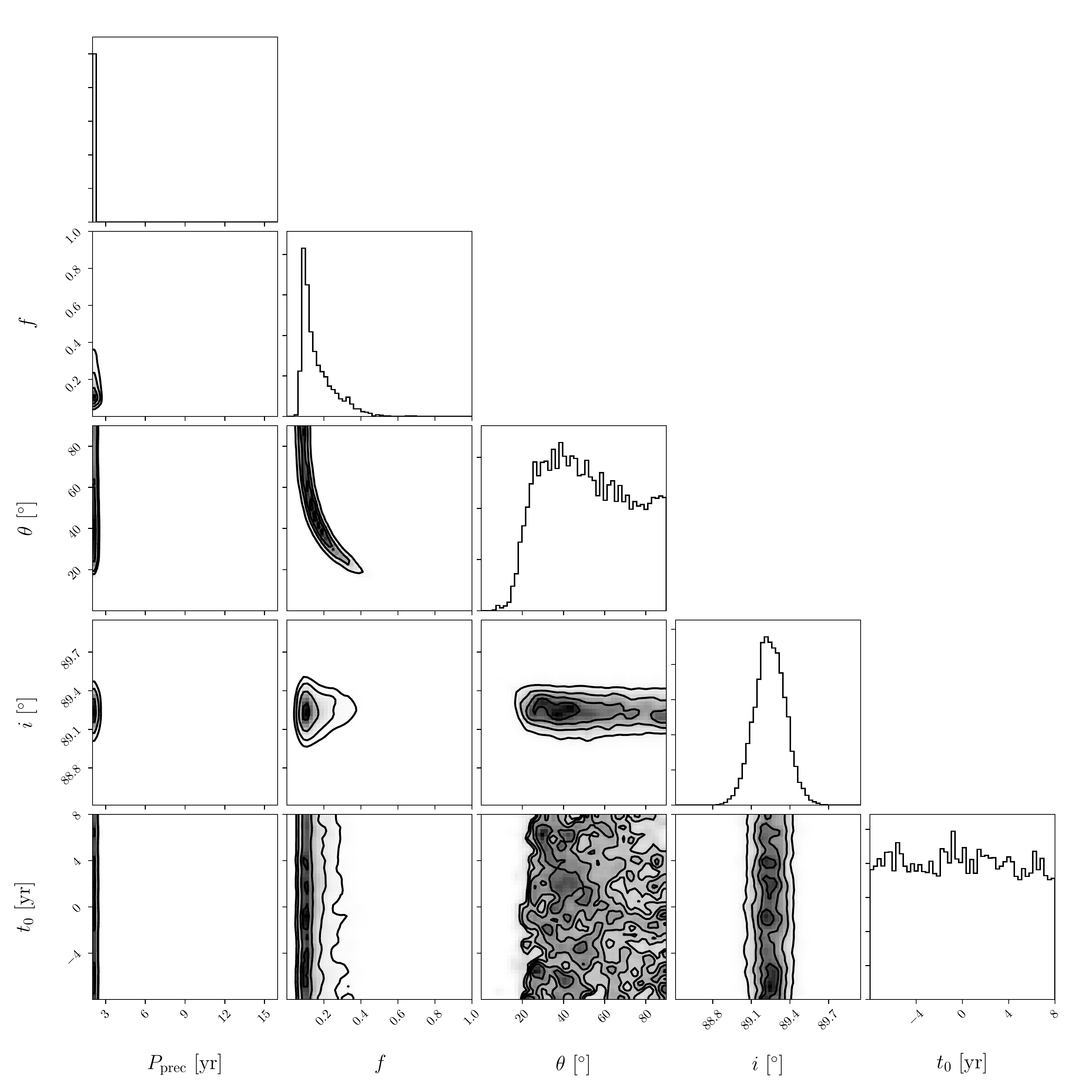}
	\caption{\small Posterior probability distributions from MCMC analysis of Kepler-39b transit depths. Contours are drawn at $0.5 \sigma$ intervals from $0.5 - 2.0 \sigma$ and the plots have been smoothed to remove noisy features.}
	\label{fig: KOI423 Data Corner}
\end{figure*}

\begin{figure*}[h]
	\centering
	\includegraphics[width=1.0\textwidth]{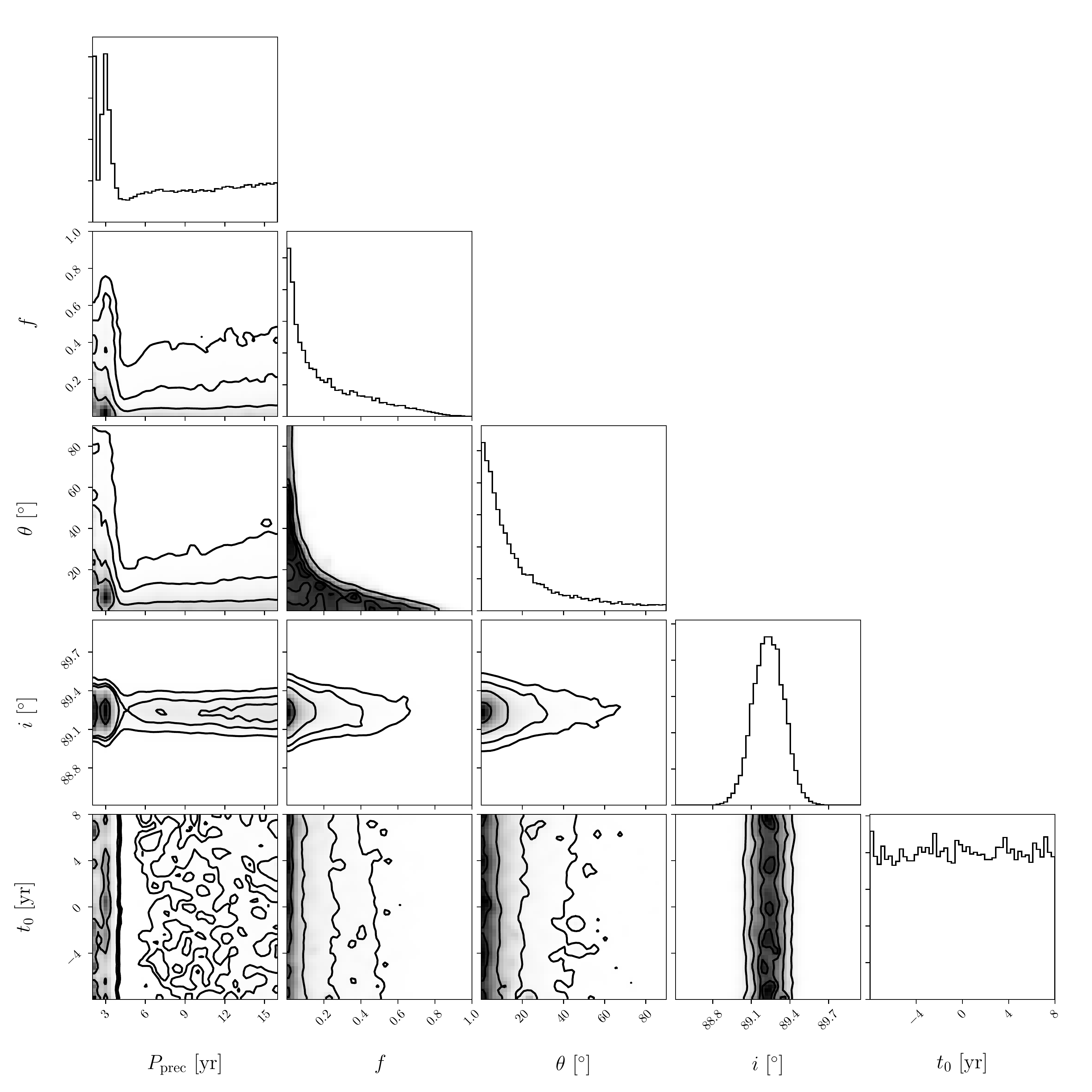}
	\caption{\small Posterior probability distributions from MCMC analysis of transit depths from Saturn-like planet transits injected into Kepler-39 photometry. Contours are drawn at $0.5 \sigma$ intervals from $0.5 - 2.0 \sigma$ and the plots have been smoothed to remove noisy features.}
	\label{fig: KOI423 Synth Corner}
\end{figure*}

Although we detected oblateness ($f = 0.13^{+0.12}_{-0.04}$) with high significance in Kepler-39b, based on the additional injection and recovery tests, this detection is likely a false positive. The posterior distributions of precession period recovered from the true and Saturn-like simulated Kepler-39 photometry (Figures \ref{fig: KOI423 Data Corner}, \ref{fig: KOI423 Synth Corner}) both show excess probability mass at low periods. The recovered 2.04 year period is exactly twice the 372.5 day \textit{Kepler} orbit and suggests an explanation.
For $i \approx 90^\circ$, the $\mathrm{T} \delta \mathrm{V}$ peaks have comparable amplitude, creating a quasi-periodic signal matching the \textit{Kepler} year. 
While we have corrected for sharp changes in transit depth at quarter boundaries caused by the spacecraft's seasonal rolls (see Section \ref{sec: crowding}), other systematics have been observed. In a study of M giant variability, \citet{2013MNRAS.436.1576B} found one quarter of their targets exhibited smooth flux variations with 372.5 day periods. The high statistical significance ($p = 0.002$, $3.09\sigma$) of the observed transit depth variations may reflect detection of this effect or another similar systematic.

The physical plausibility of such rapid precession casts additional doubt on our detection. Assuming the detected oblateness to be rotationally induced, we can calculate the required rotation rate and estimated precession period using Equations \eqref{eq: f rot} and \eqref{eq: precession period}. \citet{2015AA...575A..85B} recently updated the parameters for the Kepler-39 system. They find that the previously measured eccentricity may be spurious, so we adopt the values from the circular model for simplicity: $M_p = 19.1 \pm 1 \ M_J$ and a mean radius $R_m = \sqrt{R_\mathrm{eq} R_\mathrm{pol}} = 1.11 \pm 0.03 \ R_J$. For planets in hydrostatic equilibrium, we can apply the Darwin-Radau approximation \citep{1999ssd..book.....M}:
\begin{align}
\label{eq: darwin-radau}
\frac{J_2}{f} = -\frac{3}{10} + \frac{5}{2} \lambda - \frac{15}{8} \lambda^2 \mathrm{.}
\end{align}
Assuming, conservatively, that $\lambda = 0.4$, corresponding to a uniform density sphere, then $J_2 = 0.052$ and the required rotation period to produce the measured oblateness is $P_\mathrm{rot} = 2.7~\mathrm{hours}$. The resulting precession period is $P_\mathrm{prec} = 87~\mathrm{years}$. Under the more common assumption that $\lambda \approx 0.23$ for giant planets, $P_\mathrm{prec} \approx 150~\mathrm{years}$.
Relaxing the Darwin-Radau constraint, assuming the observed oblateness can be supported at $P_\mathrm{rot} \sim 5~\mathrm{hours}$, and that $\cos{\theta} \sim 1$, the observed precession period requires $\lambda/J_2 \sim 0.5$, considerably less than the value of 13.5 for Saturn \citep{2004AJ....128.2501W}. 

The presence of exomoons around Kepler-39b could alter the effective value of $\lambda / J_2$ and drive faster precession; Saturn's satellites reduce its calculated precession period by a factor of 4 \citep{2010ApJ...716..850C}.
The effective value of $\lambda/J_2$ is given by $(\lambda + l) / (J_2 + j)$ where
\begin{align}
l = \sum_{i} \frac{m_i}{M_p} \left( \frac{a_i}{R_\mathrm{eq}} \right)^2 \frac{P_\mathrm{rot}}{p_\mathrm{orb,~i}} \ \textrm{, and}
\\
j = \frac{1}{2} \sum_{i} \frac{m_i}{M_p} \left( \frac{a_i}{R_\mathrm{eq}} \right)^2 \frac{\sin{(\theta - I_i)}}{\sin{\theta}} \textrm{,}
\end{align}
where $m_i$, $a_i$, $p_\mathrm{orb,~i}$, and $I_i$ are the satellite's mass, semimajor axis, orbital period and inclination relative to the planet's equator \citep{2004AJ....128.2501W}. 
Prograde satellites on circular orbits are only stable with a semimajor axis $a \lesssim 0.4895~R_\mathrm{Hill}$ \citep{2006MNRAS.373.1227D, 2008ApJ...686..741S}.
Under the generous (and inconsistent) assumptions that $\lambda = 0.1334$ (corresponding to $J_2 \approx 0$ in the Darwin-Radau relationship), $J_2 = 0.052$, and that the satellite has $a = 0.4895~R_\mathrm{Hill}$, then the required reduction in $\lambda / J_2$ is achieved at $m \approx 5 \ M_\oplus$. 
By comparison, Jupiter and Saturn's largest satellites, Ganymede and Titan, have masses of $0.025 \ M_\oplus$ and $0.022 \ M_\oplus$, respectively \citep{Showman77, 2006AJ....132.2520J}, and each orbit at ${\sim} 0.02$ times the Hill radius of their respective hosts. Modeling satellite formation around gas giants, \citet{2006Natur.441..834C} find that competing processes naturally limit satellite systems to a total mass of ${\sim}10^{-4}~M_p$, or ${\sim}0.6~M_\oplus$ for Kepler-39b. A massive satellite of Kepler-39b, therefore, seems to be an unlikely explanation for the observed precession period.

If our detection is indeed a false positive, these results are consistent with the findings of \citet{2014ApJ...796...67Z}. Using their assumptions and measured values of $f_\perp = 0.22 \pm 0.11$ and $\theta_\perp = -40^\circ$, we obtain an expected precession period of ${\sim} 100$ years, much longer than could be detected from 4 years of \textit{Kepler} observations. 
Updated measurements of the Kepler-39 system \citep{2015AA...575A..85B} have reduced the age estimate from $5.1 \pm 1.5 \ \mathrm{Gyr}$ to $1.0^{+0.9}_{-0.7} \ \mathrm{Gyr}$ (or $2.1^{+0.8}_{-0.9} \ \mathrm{Gyr}$ if the observed eccentricity is real), greatly decreasing the time available to de-spin the planet and bolstering the claim that the rapid rotation could be primordial.

\subsection{Kepler-427b (KOI 192.01)}
We detected variation in the transit depths of Kepler-427b consistent with moderate oblateness ($f = 0.19^{+0.32}_{-0.16}$) at a significance level of $1.65\sigma$. In contrast to Kepler-39b, we do not believe this to be a data artifact.
While the simulated null case does show an excess of probability density at short precession periods, there is no indication of this signal in the recovered distributions for the Saturn-like case or the real data. Furthermore, the solutions found for the real data and Saturn-like planet are not located in the same region of parameter space as the signal found for the null case.

If the observed signal is caused by spin precession, then armed with the measured oblateness and precession period, and assuming the Darwin-Radau approximation, we can infer a range of values for $\lambda$, $J_2$ and $P_\mathrm{rot}$. Adopting $M_p = 0.29 \pm 0.09 \ M_J$ and $R_m = 1.23 \pm 0.21 \ R_J$ from \citet{2014AA...572A..93H}, we find $\lambda \simeq 0.16^{+0.08}_{-0.02}$, $J_2 \simeq 9000^{+4000}_{-3000} \times 10^{-6}$, and $P_\mathrm{rot} \simeq 15^{+22}_{-2} \ \mathrm{hours}$. 
By comparison, Jupiter has $J_2  = 14695.62 \pm 0.29 \times 10^{-6}$ and a rotation rate of $10 \ \mathrm{hours}$.\footnote{\url{http://ssd.jpl.nasa.gov/?gravity_fields_op}} The detailed interior structure of Jupiter is still uncertain, but common model-derived values for the normalized moment of inertia are $\lambda = 0.26387 - 0.26394$ \citep{2016ApJ...820...80H}. 
Saturn, with nearly the same mass as Kepler-427b, has $J_2 = 16290.71 \pm 0.27 \times 10^{-6}$ \citep{2006AJ....132.2520J} and a rotation period of $10.7 \ \mathrm{hours}$. \citet{1989Icar...78..102H} present an interior model for Saturn with $\lambda = 0.22037$ while \citet{2004AJ....128.2501W} propose a dynamical origin for Saturn's obliquity which requires $0.2233 < \lambda < 0.2452$. Most of the range of our inferred values is outside that expected from Solar System gas giants, but there is some overlap.

The rotation period we measure is significantly shorter than the orbital period of the planet. Using Equation \eqref{eq: spin down time}, assuming an initial rotation period of $9 \ \mathrm{hours}$, near the breakup rotation rate, and an optimistic tidal dissipation factor, $Q = 10^{6.5}$, we calculate that only ${\sim} 10~\mathrm{Myr}$ are required to spin down Kepler-427b to the rotation periods inferred by our measurement. This is much shorter than the estimated system age of $7 \pm 4 \ \mathrm{Gyr}$, an apparent contradiction. Equation \eqref{eq: spin down time} is highly approximate, and is derived assuming a circular orbit and a planet with its spin axis aligned to the orbit normal. 
Planets on eccentric orbits do not directly synchronize, instead entering a quasi-synchronous spin state while they retain significant eccentricity \citep{2004ApJ...610..464D}. \citet{2014AA...572A..93H} were not able to significantly measure the eccentricity of Kepler-427b, providing only a $99\%$ upper bound of $e < 0.57$.
 Kepler-427b lies in the ``period valley''\textemdash the sparsely populated region between hot Jupiters and more distant gas giant planets ($10 \lesssim P \lesssim 85 \ \mathrm{days}$)\textemdash making tidal circularization a precarious assumption. \citeauthor{2014AA...572A..93H} suggest that follow up measurements could more precisely constrain the orbital eccentricity, stellar age and radius, and the planetary radius, potentially resolving the spin-down time problem.

\subsection{Other Causes of Transit Variations}
Changes in the transit geometry can also alter the measured transit depth. To confirm our $90.1\%$ confidence level we examined several possible sources of false positives.

\subsubsection{Nodal Precession}
If Kepler-427b is undergoing sufficiently rapid nodal precession\textemdash the precession of the orbital plane\textemdash the impact parameter of the transit would change during our observations. In conjunction with limb darkening, this would change the measured transit depth over time. The nodal precession rate is given by 
\begin{align}
\omega_\mathrm{nodal} = - \frac{3}{2} \frac{R_*^2}{a^2 (1 - e^2)^2} J_{2,*} \frac{2 \pi}{P_\mathrm{orb}} \cos{\psi} \text{,}
\end{align}
where $\psi$ is the angle between the orbit normal and the spin axis of the star and $J_{2,*}$ is the star's zonal quadrupole moment  \citep{1999ssd..book.....M, 2013ApJ...774...53B}. For the Kepler-427 system, $R_* = 1.35 R_\odot$, and $a = 0.091~\mathrm{AU}$ \citep{2014AA...572A..93H}. For the stellar quadrupole moment, we adopted the Sun-like value $J_{2,*} \sim10^{-7}$ \citep{1981ApJ...246..985U, 2004SoPh..222..191M}. Assuming, conservatively, that $\cos{\psi} \approx 1$ and that the eccentricity is near the upper bound, $e = 0.57$, we obtain $| w_\mathrm{nodal} | \sim 10^{-15}~\mathrm{rad}~\mathrm{s}^{-1}$. This is a nodal precession period of ${\sim}40~\mathrm{Myr}$, too slow to account for the observed changes in transit depth.

\subsubsection{Apsidal Precession}
If Kepler-427b is on an eccentric orbit, then apsidal precession, or periastron precession, would change the star-planet distance during the transit over time. If the transit is not observed edge-on, this would cause the transit impact parameter, and observed transit depth, to change over time. \citet{2009ApJ...698.1778R} calculated the expected apsidal precession rate for a hot Jupiter. Following their formulation, we find the dominant term is precession driven by the planet's rotational bulge:
\begin{multline}
\omega_\mathrm{rot,~p} = \frac{k_{2,~p}}{2} \left( \frac{R_p}{a} \right)^5 \left( \frac{2 \pi}{P_\mathrm{rot}} \right)^2 
\\
\times \frac{a^3}{G M_p (1 - e^2)^2} \left( \frac{2 \pi}{P_\mathrm{orb}} \right)^2 \text{,}
\end{multline}
where $k_{2,~p}$ is the planet's Love number. Assuming $e = 0.57$, $P_\mathrm{rot} = 15~\mathrm{hours}$, and a Saturn-like Love number, $k_{2,~p} \approx 0.4$ \citep{2017Icar..281..286L}, we calculate $\omega_\mathrm{rot,~p} = 3.2 \times 10^{-11}~\mathrm{s}^{-1}$. To include the remaining terms from \citeauthor{2009ApJ...698.1778R}, we assumed a stellar rotation period of ${\sim}10~\mathrm{days}$, appropriate for Sun-like stars with ages of ${\sim}1~\mathrm{Gyr}$ \citep{2014ApJ...790L..23D}, and consistent with $v \sin{i_*}  = 3 \pm 1 ~\mathrm{km}~\mathrm{s}^{-1}$, measured by \citet{2014AA...572A..93H}. Following \citeauthor{2009ApJ...698.1778R}, we adopted a stellar Love number $k_{2,*} = 0.03$.
 Under these assumptions, the total precession induced is $\omega_\mathrm{apsidal} = 4.5 \times 10^{-11} ~\mathrm{rad}~\mathrm{s}^{-1}$. This includes, in order of diminishing importance, the effects of the rotational bulge of the planet, the tidal bulge on the planet, general relativity, the rotational bulge of the star, and the tidal bulge on the star.
  The corresponding apsidal precession period of $4.4~\mathrm{kyr}$ is too long to explain the observed change in transit depth.

\subsubsection{Three Body Interactions}
The preceding calculations only considered a two-body system. Dynamical interactions with another planet in the Kepler-427 system could drive faster precession or create secular variation in the eccentricity or inclination of Kepler-427b's orbit. While we cannot definitively exclude this possibility, we found no evidence of timing variations in the transits of Kepler-427b, consistent with the TTV survey of the entire Kepler data set undertaken by \citet{2016ApJS..225....9H}.
 Additionally, no evidence of another companion is reported in the HARPS-N spectrograph observations conducted by \citet{2014AA...572A..93H}.

Given the expected long precession periods and lack of evidence for strong interactions with another body, we conclude that changes in orbital geometry are not the cause of the observed transit depth variations, and that the $90.1\%$ confidence estimate is reliable.

\subsection{Conclusion}
We have presented the first attempt at detecting the oblateness of an exoplanet through changes in transit depth caused by spin precession. We examined two planets in detail. While we were unable to detect the oblateness of Kepler-39b, this is broadly consistent with the findings of \citet{2014ApJ...796...67Z} given the expected long precession period. 
We find transit depth changes consistent with an oblateness comparable to Solar System gas giants for Kepler-427b, but with a significance of only $90.1\%$ ($1.65 \sigma$). Kepler-427b is a warm Saturn in the period-valley ($P \approx 10{-}85~\mathrm{days}$), a class of planets with an unclear formation mechanism \citep{2016A&A...587A..64S}. Confirming and improving this oblateness detection and further constraining the bulk properties of Kepler-427b would illuminate the planet's internal structure, possibly providing insight into period-valley giant planet formation, making the Kepler-427 system an attractive target for followup observations.

Current and near-future missions, such as K2 \citep{2014PASP..126..398H} using \textit{Kepler} and the \textit{Transiting Exoplanet Survey Satellite} \citep{2015JATIS...1a4003R}, probably do not have the long time baseline required to detect these transit depth variations. TESS, however, is likely to provide a wealth of more easily characterizable targets in the period range where these effects are measurable, allowing for long-term followup using other instruments. Additionally, TESS's high quality short-cadence photometry will provide a rich dataset for the method attempted by \citet{2010ApJ...709.1219C} and \citet{2014ApJ...796...67Z}. Efforts combining both methods promise to greatly expand our understanding of gas giant structure and formation.

\acknowledgements
We thank Zachory K. Berta-Thompson for patient instruction in the careful interpretation of \textit{Kepler} photometry and Joshua Winn for insightful comments which substantially improved the manuscript.
We also extend our gratitude to Margaret Pan and Jennifer Burt for helpful discussions, as well as Niraj Inamdar, Alexandria Gonzalez, and the MIT Exoplanet discussion group.
Finally, we thank Joshua Carter whose initial work inspired this paper and the anonymous referee for their feedback.

\software{emcee \citep{emcee}}

\bibliographystyle{plainnat}
\bibliography{OblatePlanets}

\end{document}